\documentclass[journal,onecolumn,compsoc]{IEEEtran}
\usepackage[T1]{fontenc}
\usepackage{footnote}
\usepackage{url}
\usepackage{cite}
\usepackage{paralist}
\usepackage{array}
\usepackage{supertabular}
\usepackage{booktabs}
\usepackage{multirow}
\usepackage{enumitem}
\usepackage{xspace}
\usepackage{balance}
\usepackage{tikz}

\usepackage{amssymb}

\newcounter{guideline}
\newenvironment{guideline}{\refstepcounter{guideline}
	\vspace{2pt}
	\begin{compactitem}
		\item[(G\theguideline)] \it }
	{ \end{compactitem}\smallskip}

\newcounter{practice}
\newenvironment{practice}{\refstepcounter{practice}
	\vspace{2pt}
	\begin{compactitem}
		\item[(P\thepractice)] \it }
	{ \end{compactitem}\smallskip}

\newcounter{challenge}
\newcommand{\challenge}{\refstepcounter{challenge}}

\newcounter{finding}
\newenvironment{finding}{\refstepcounter{finding}
	\vspace{2pt}
	\begin{compactitem}
		\item[F\thefinding:] \it }
	{ \end{compactitem}\smallskip}

\newenvironment{widequotation}{\list{}{\listparindent 0em \itemindent\listparindent
		\leftmargin 15pt	\rightmargin 2pt \parsep 0pt plus 1pt}\item\relax}
{\endlist}
\def\signed#1{{\leavevmode\unskip\nobreak\hfil\penalty50\hskip2em
		\hbox{}\nobreak\hfil\raise-1pt\hbox{#1}%
		\parfillskip=0pt \finalhyphendemerits=0 \endgraf}}
\newsavebox\mybox
\newenvironment{aquote}[1]
{\savebox\mybox{(#1)}\begin{widequotation}\itshape``\ignorespaces}
	{\unskip''\signed{\usebox\mybox}\end{widequotation}}

\newcommand{\jx}{a product owner\xspace}
\newcommand{\Jx}{A product owner\xspace}
\newcommand{\vx}{a requirements manager\xspace}
\newcommand{\Vx}{A requirements manager\xspace}
\newcommand{\tx}{a chief architect\xspace}
\newcommand{\Tx}{A chief architect\xspace}
\newcommand{\ndx}{a logical architect\xspace}
\newcommand{\Ndx}{A logical architect\xspace}
\newcommand{\ru}{a process manager\xspace}
\newcommand{\Ru}{A process manager\xspace}
\newcommand{\ao}{a team leader for functional development\xspace}
\newcommand{\Ao}{A team leader for functional development\xspace}
\newcommand{\ra}{a Scrum master\xspace}
\newcommand{\Ra}{A Scrum master\xspace}

\newcommand{\Mux}{A product owner\xspace}

\newcommand{\IntC}{A system architect from Company C\xspace}
\newcommand{\intB}{a software architect from Company B\xspace}

\newcommand{\jxR}{Product owner\xspace}
\newcommand{\vxR}{Requirements manager\xspace}
\newcommand{\txR}{Chief architect\xspace}
\newcommand{\ndxR}{Logical architect\xspace}
\newcommand{\ruR}{Process manager\xspace}
\newcommand{\aoR}{Team leader\xspace}
\newcommand{\raR}{Scrum master\xspace}
\newcommand{\muxR}{Product owner\xspace}

\begin{document}
\title{Boundary Objects and their Use in Agile Systems Engineering}

\author{	
	{Rebekka Wohlrab $^{(1,2)}$, Patrizio Pelliccione $^{(1, 3)}$, Eric Knauss $^{(1)}$, Mats Larsson $^{(2)}$}\\
	$^{(1)}$ Department of Computer Science and Engineering, Chalmers | University of Gothenburg, Gothenburg, Sweden \\
	$^{(2)}$ Systemite AB, Gothenburg, Sweden \\
	$^{(3)}$ University of L'Aquila, Italy
}


\maketitle

\begin{abstract}
	Agile methods are increasingly introduced in automotive companies in the attempt to become more efficient and flexible in the system development.
	The adoption of agile practices influences communication between stakeholders, but also makes companies rethink the management of artifacts and documentation like requirements, safety compliance documents, and architecture models.
	Practitioners aim to reduce irrelevant documentation, but face a lack of guidance to determine what artifacts are needed and how they should be managed.
	This paper presents artifacts, challenges, guidelines, and practices for the continuous management of systems engineering artifacts in automotive based on a theoretical and empirical understanding of the topic.	
	In collaboration with 53 practitioners from six automotive companies, we conducted a design-science study involving interviews, a questionnaire, focus groups, and practical data analysis of a systems engineering tool.
	The guidelines suggest the distinction between artifacts that are shared among different actors in a company (boundary objects) and those that are used within a team (locally relevant artifacts).
	We propose an analysis approach to identify boundary objects and three practices to manage systems engineering artifacts in industry.
	
	\begin{tikzpicture}[overlay]
	\node[align=center, draw, fill=white, thick, text width=18cm, anchor=north] (b) at (8,6.5)
	{\baselineskip=10pt  This is the accepted version of the following article: Boundary Objects and their Use in Agile Systems Engineering, which has been published in the Journal of Software: Evolution and Process.  https://doi.org/10.1002/smr.2166
		
		This article may be used for non-commercial purposes in accordance with the Wiley Self-Archiving Policy [http://www.wileyauthors.com/self-archiving].};
	

	\end{tikzpicture}
\end{abstract}

\section{Introduction}
Continuous software engineering and agile methods have been researched for more than two decades and are increasingly used in industry.
Agile approaches originate from the area of software development and bring a number of reported and desired benefits, e.g., better customer collaboration, increased productivity, and better product quality~\cite{Dyba2008}.
Hoping for similar benefits and successes as in the field of software development, there have been attempts to apply agile methods also in systems engineering and product development contexts~\cite{Houston2014}.
The term ``agile systems engineering'' has been used for more than a decade (e.g.,~\cite{Turner2007,Haberfellner2005}).
In this paper, we follow the notion of agile SYSTEMS ENGINEERING, i.e., focusing on agility in the process of engineering systems rather than on agility in the systems themselves~\cite{Haberfellner2005}.
We use the term ``large-scale agile development'' to refer to agile development efforts with a large number of actors, systems and inter-dependencies, and involving more than two teams~\cite{Dingsoyr2017a}.
For instance, agile methods are adopted in automotive organizations to increase the level of flexibility to change and stay competitive to other cost-effective companies~\cite{Ebert2017}.
However, the automotive domain comes with several challenges: The ever-growing system complexity, safety criticality, a plethora of variants, and the issue of knowledge management over decades in large organizations~\cite{DAmbrosio2017,Ebert2017}.

Knowledge is often manifested in artifacts like architecture descriptions, requirements, or test cases.
In this paper, we use the term \emph{systems engineering artifacts} to refer to artifacts as entities capturing information concerned with systems engineering.
When applying agile approaches, practitioners aim to reduce irrelevant documentation and improve knowledge management practices~\cite{Pareto2010,Hummel2013}.

A practical challenge is to find the `right' amount of documentation in agile contexts~\cite{Ruping2003}.
Especially in large-scale agile development, the need for more research on knowledge sharing and inter-team coordination has been identified to provide better advice for practitioners~\cite{Dingsoyr2017}.
Automotive companies require better guidance to identify what documentation is needed and understand how it can be managed in large-scale agile development contexts~\cite{Kajko-Mattsson2008}.

Systems engineering in large-scale organizations consisting of multiple groups is intrinsically connected to social aspects~\cite{Scacchi1984}.
As the topic of knowledge management is strongly related to the organizational structure with a network of social groups, we base parts of our work on theoretical frameworks from the field of sociology.
Concretely, we analyze and describe how actor-network theory can be leveraged to conceive how teams interact with each other in an automotive organization.
Particularly, we focus on the concept of ``boundary objects''~\cite{Star1989}, artifacts used to collaborate across boundaries between several teams.
We relied on this concept to explore what artifacts are crucial for multiple groups and how these artifacts can be managed best.
The theoretical foundation allows us to strengthen the paper by allowing us to understand the potential of boundary objects from a sociological perspective.
We used the theoretical understanding to develop guidelines for the management of systems engineering artifacts and contribute to the body of knowledge of agile information management.

This paper presents artifacts, challenges, guidelines, and practices for the continuous management of systems engineering artifacts in automotive based on a theoretical and empirical understanding of the topic.	
This paper provides a significant extension of our previous contribution~\cite{Wohlrab2018}:
We greatly extend the description of the theoretical foundations of actor-network theory and boundary objects in which this work is to be understood.
Moreover, we investigate an additional research question about practical support to manage systems engineering artifacts with a systems engineering tool.
To answer this question, we analyzed collaboratively managed development data stored in the systems engineering tool SystemWeaver\footnote{\url{http://systemweaver.se/}}, developed an approach to identify boundary object candidates, and created practices that are in line with our guidelines.
Finally, we provide more details and a more elaborate discussion of our findings, related work, and the implications of the paper for researchers and practitioners.
Concretely, we investigate the following research questions:

\newcommand{\rqArtifacts}{What are practices to manage artifacts in agile automotive systems engineering?}
\newcommand{\rqStrategies}{What guidelines could support the management of systems engineering artifacts in agile automotive contexts?}
\newcommand{\rqChallenges}{What practical challenges exist with managing systems engineering artifacts in agile automotive contexts?}
\newcommand{\rqPractical}{How can the management of systems engineering artifacts in agile automotive contexts be supported by a systems engineering tool?}

\textbf{RQ1:} \rqArtifacts

\textbf{RQ2:} \rqChallenges

\textbf{RQ3:} \rqPractical

The understanding of systems engineering artifacts based on these research questions is the starting point for the iterative creation of guidelines to manage systems engineering artifacts in practice.

Our research method was a design-science study~\cite{Hevner2004} in which we developed and evaluated guidelines for the continuous management of systems engineering artifacts.
We collaborated with six automotive companies attempting to increase their agility and minimize irrelevant documentation.
In the study, we reviewed related work and conducted a case study to explore artifacts, practices, and challenges.
In particular, we focused on concepts from the domain of sociology to understand the applicability of boundary objects to our domain.
Several iterations were used to create and evaluate guidelines using focus groups, interviews, and a questionnaire.
Additionally, we analyzed data from an automotive OEM, developed an analysis approach for boundary object candidates, and suggest practices to manage artifacts using a tool.

Four automotive OEMs participated in our study, as well as an automotive supplier, and a supplier of an information management tool used in the automotive domain.
The individuals from these companies were 53 experts with diverse backgrounds and 31 anonymous questionnaire respondents.

\noindent \textbf{Contributions:} Our contributions are
\begin{enumerate}[noitemsep,topsep=0pt]
	\item an analysis of systems engineering artifacts and practices to manage them,
	\item a catalog of challenges,
	\item a conceptualization of boundary objects in the field of systems engineering,
	\item guidelines to manage systems engineering artifacts,
	\item an analysis of how systems engineering artifacts can be supported by a tool
\end{enumerate}

Our findings indicate that in automotive organizations, it is recommendable to distinguish between artifacts used to cross boundaries between groups of actors and those used within a team.
It is especially crucial to understand boundary objects and support a lightweight approach to create them for early impact analysis and decision making.
Artifacts used within a team can be managed more flexibly by the team and be created as late as desired.
Practical challenges include the degradation of artifacts and collaboration across multiple locations.
Our guidelines address these challenges and can be supported by a practical systems engineering tool.
We acknowledge the need for future studies to analyze our findings' generalizability, but expect the guidelines to be applicable to other systems engineering contexts.
Our research was focused on the automotive domain, however, related work has shown that the reported challenges are also prominent in other domains~\cite{Dingsoyr2017}.

The remainder of this paper is structured as follows:
Section~\ref{sec:RelatedWork} presents related work.
Section~\ref{sec:Theoretical} elaborates on the theoretical foundation of boundary objects in systems engineering.
Section~\ref{sec:ResearchMethod} describes our research method.
Our findings are presented in Section~\ref{sec:Finding} and discussed at the end of each of its subsections.
In Section~\ref{sec:SystemWeaver}, we elaborate the support of managing boundary objects with a systems engineering tool.
We give an overview of our findings in Table~\ref{tab:Findings1}.
In Section~\ref{sec:Discussion}, we discuss the overview of our findings, describe our conclusions, and present future work.

\section{Related Work} \label{sec:RelatedWork}
This section presents related work concerned with the role of communication in agile approaches, artifacts in large-scale agile contexts, boundary objects, and characteristics of the automotive domain.

\paragraph*{ \em Communication in Agile Approaches}
Previous empirical studies have focused on communication aspects in agile contexts.
Due to the stronger focus on face-to-face conversations and direct knowledge exchange, Petersen and Wohlin found that intra-team communication improved~\cite{Petersen2010}.
However, they concluded that coordination between teams is generally considered a challenge.
For this reason, the need to find different communication modes for distributed agile teams has been proclaimed.
Different communication modes are even more ineluctable when knowledge needs to be retained for maintenance~\cite{Hummel2013}.
Another study concluded that knowledge sharing in an agile team can be facilitated, but the need to find agile ways of documentation remains~\cite{Lagerberg2013}.

\paragraph*{ \em Artifacts in Large-Scale Agile Contexts}
Artifacts play a central role for various software and systems engineering activities, as artifacts are the forms in which knowledge is manifested.
However, there exists no common understanding in the community with respect to what an artifact is and how it should be defined.
Attempts have been made to create artifacts models and identify artifacts, especially in requirements engineering, but also in distributed agile projects~\cite{Kuhrmann2013,Mendez2010}.
In a recent paper, M\'endez Fern\'andez et al.~aimed to contribute to a common understanding of artifacts by establishing fundamental concepts.
In this paper, we follow their definition of an artifact as ``a self-contained work result, having a context-specific purpose and constituting a physical representation, a syntactic structure and a semantic content, forming three levels of perception''~\cite{Mendez2018}.

Bass raised an inventory of artifacts in large-scale agile development and identified 25 artifacts used for a variety of development, release, and governance activities~\cite{Bass2016}.
However, the conclusion they arrived at was that ``there are no agile ceremonies specifically designed to create and refine any of these artifacts'', and this complicates the coordination of collaborating teams~\cite{Bass2016}.
The research presented in this paper addresses this issue by presenting guidelines to manage artifacts, with a focus on inter-team collaboration.

Another classification of typical artifacts in agile practices builds on the categories of management, specification, design, migration, usage, test, and operations~\cite{Ruping2003}.
However, the elicitation of documentation requirements and the identification of an appropriate amount of documentation were found to be complicating factors.
For this purpose, related work has pointed to the need for more ``research on methods and tools that facilitate agile documentation''~\cite{Voigt2016}.

Another classification of models that we considered in this study are descriptive vs.~prescriptive artifacts~\cite{Heldal2016}.
Descriptive artifacts ``describe a subject that already exists'', which can be implicit in the modeler's mind (an idea of the future architecture) or explicit (e.g., capturing an already running system). 
Prescriptive artifacts are leveraged to create another subject that does not exist yet, e.g., models used to generate code or other artifacts.
We found this categorization to be valuable also for agile contexts, as the purpose of artifacts greatly impacts how they are used and managed.

\paragraph*{ \em Boundary Objects}
Organizations performing systems engineering can be conceived as a network of actors that interact using (potentially shared) resources and with multiple goals.
This conception is influenced by coordination theory~\cite{Malone1994} and actor-network theory~\cite{Callon1986,Latour1987}.
Boundary objects are a means of enabling collaboration between different groups of actors, as we describe more detailedly in Section~\ref{sec:Theoretical}.
Boundary objects in software engineering have been analyzed by several studies.
For instance, Pareto et al.~\cite{Pareto2010} examined architectural descriptions as boundary objects in agile systems engineering and specified requirements for their creation~\cite{Pareto2010}.
Another study focused on boundary objects in distributed agile user-centered design~\cite{Blomkvist2015}, concluding that boundary objects are facilitators for communication in that domain.
The production process was in the focus of another study, analyzing how engineers, assemblers, and technicians communicate across boundaries~\cite{Bechky2003}.
However, to the best of our knowledge, there exists no study focusing on boundary objects in the context of agile systems engineering.

\paragraph*{ \em Challenges in the Automotive Domain}
The automotive industry is the domain in which our study is situated.
It comes with several challenges, as it involves the development of complex systems in collaboration with thousands of stakeholders from a plenitude of disciplines (mechanical engineering, software engineering, economics, electronics).
Moreover, the automotive domain requires a variety of quality attributes to be fulfilled in distributed development environments with short cycle times~\cite{Ebert2017}.
It is a domain in which knowledge management~\cite{DAmbrosio2017} is of crucial importance, which requires ways to manage traceability~\cite{Wohlrab2016,Maro2018} and scattered information~\cite{Pretschner2007}.
Challenges can be observed when designing the software architecture of automotive systems, as practitioners struggle with the creation of meaningful documentation that can be shared across organizational boundaries~\cite{Pelliccione2017}.
Practitioners require support for agile architecting and communication across team borders~\cite{Pelliccione2017}.
Also in the domain of requirements engineering, automotive companies struggle with aligning distributed teams and finding agile mechanisms to communicate effectively~\cite{Eliasson2015RE,Wohlrab2018REFSQ}.
At the same time, several standards play a substantial role in automotive, for instance, ensuring safety (e.g., ISO 26262~\cite{ISO26262}) or quality of software processes (e.g., Automotive SPICE~\cite{Automotive2017}).
Diebold et al.~\cite{Diebold2017} stated that the combination of agile methods with Automotive SPICE is possible, but that more guidance is still needed, e.g., with respect to architecture documentation.
Our research works towards more practical guidance by suggesting guidelines to manage systems engineering artifacts.

\section{Theoretical Foundation: Boundary Objects in Systems Engineering} \label{sec:Theoretical}

\begin{figure}
	\centering
	\includegraphics[width=0.8\linewidth]{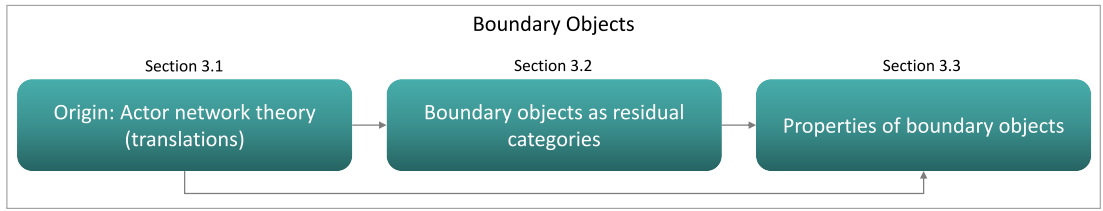}
	\caption{An overview of the presented theoretical concepts}
	\label{fig:theory}
\end{figure}

This section introduces the theoretical foundation of this paper to better understand the use of systems engineering artifacts in an agile automotive organization.
The concept of \emph{boundary objects} is the key concept we focus on in this paper.
It is commonly defined as follows:
\begin{quotation}
	\emph{Boundary objects} are objects which are both plastic enough to adapt to local needs and the constraints of the several parties employing them, yet robust enough to maintain a common identity across sites.~\cite{Star1989}
\end{quotation}

In the following, we present the underlying theoretical framework to elaborate on the properties of boundary objects.
An overview of the presented concepts is shown in Figure~\ref{fig:theory}.
In Section~\ref{sec:Theoretical:ActorNetwork}, we describe the origin of boundary objects in the field of actor network theory and the translations between social groups.
Section~\ref{sec:Theoretical:ResidualCategories} presents how boundary objects can be understood as residual categories in the process of classification.
Finally, based on this background, we elaborate on properties of boundary objects in Section~\ref{sec:Theoretical:Properties}.

\subsection{Origin of Boundary Objects: Actor-Network Theory} \label{sec:Theoretical:ActorNetwork}
To analyze the concept of boundary objects in a systems engineering context, it is essential to first shed light on the interdisciplinary groups of people involved in these organizations---the ``social groups'', or ``parties'', as in the definition above.
When approaching an organization, it becomes evident that there are several social groups involved in visible or invisible ways.

In an organization performing systems engineering, as in other social contexts, groups can have separate concerns and interests, and need to be aligned to collaborate effectively.
This form of reconciliation can be understood in the context of actor-network theory, initially developed by Callon~\cite{Callon1986} and Latour~\cite{Latour1987}.
Actor-network theory focuses on heterogeneous networks of human and non-human actors, organizations, and standards.
In order to align participants' interests, the process of \emph{translation} is used to bring actors into a relationship and translate between their concerns.
A successful translation eventually establishes one of the actors as a ``spokesman'' that represents the unified concerns of a group of actors (``allies'')~\cite{Callon1986}.
Translation requires negotiations and adjustments, and is complicated by the fact that translations are often initiated in an ``$n$-way nature'', with simultaneous translations initiated by all involved actors~\cite{Star1989}.
Despite the focus on aligning participants' interests, translations can also be seen from the viewpoint of knowledge management: Information needs to be conveyed to other participants, so that goals can be aligned and knowledge about a system or context can be successfully managed~\cite{Abraham2017}.
In Section~\ref{sec:RQArtifacts}, we explore what translations occur in systems engineering contexts and how they are supported by the use of artifacts.

Boundary objects originate in scenarios in which multiple translations occur in parallel to achieve common interests and concerns~\cite{Star1989}.
Star and Griesemer~\cite{Star1989} empirically examined a successful example of a multidisciplinary case that underwent several translations.
To achieve translations between multiple social groups, boundary objects were used to optimize the communication between them, while allowing for autonomy.
Boundary objects act as communication facilitators that leave freedom for diversity and are adjustable to the context of a local group:
\begin{quotation}
	\emph{Boundary objects} [...] have different meanings in different social worlds but their structure is common enough to more than one world to make them recognizable, a means of translation. The creation and management of boundary objects is a key process in developing and maintaining coherence across intersecting social worlds.~\cite{Star1989}
\end{quotation}

\subsection{Boundary Objects as Residual Categories} \label{sec:Theoretical:ResidualCategories}
Translations among several actors are often observable in how information is categorized and classified: When aligning different actors' concerns, one tries to find common ways of working and managing information in different categories.
Humans like to find structures and categories for information, which often results in standards, but also in the need to capture residual categories.
Residual categories include information that does not fit to the standard categories, but requires to broaden the scope and go outside of the structures of a social group.

Boundary objects become visible when information is manifested in an information system, with conventions of representation and people who use this information.
Objects in information systems ``can inhabit multiple contexts at once, and have both local and shared meaning''~\cite{Bowker1999}.
Bowker and Star~\cite{Bowker1999} elaborated on how each community of practice (or social world) establishes its own standards and categories.
For example, a team of software developers possesses knowledge regarding what information is related to a specific system function, what type a component has, or what requirements are related to safety.
Similarly, all groups of actors categorize and classify information in multiple ways (e.g., in relation to other objects, time, ownership, structure of the final system, ...).

These classifications can be on different levels of granularity or not exist at all in certain groups of actors.
Mechanisms need to be found to deal with classification differences when trying to align development practices and find common ways of working.
This often results in ``residual categories'' to capture objects that do not fit to the existing classification.
It is when the classifications of groups or communities of practice collide that ''boundary objects arise over time from durable cooperation among communities of practice''~\cite{Bowker1999}.
The establishment of boundary objects as common objects can overcome the clash of classifications of several communities of practice and allow them to collaborate effectively.

To understand the concept of residual categories in the context of systems engineering artifacts, one can consider the breakdown and allocation of high-level requirements to teams that are responsible for a component.
An intuitive approach could be that each requirement should be implemented by one component team that is responsible for it.
However, this does not work for cross-cutting requirements (e.g., performance requirements), affecting more than one component or impacting the system-level architecture.
In these cases, alternative solutions have to be created, allowing different ways to manage the residual category of specific cross-cutting requirements (here: performance requirements).
As a consequence, boundary objects might emerge.
\subsection{Properties of Boundary Objects} \label{sec:Theoretical:Properties}
There are a number of properties associated to boundary objects, some of which we describe in this section.
In the initial paper on boundary objects, Star and Griesemer describe them as ``weakly structured in common use, and become strongly structured in individual-site use. These objects may be abstract or concrete''~\cite{Star1989}.
When comparing definitions of boundary objects~\cite{Abraham2013}, it becomes evident that there exist two main characteristics: (1) Interpretive flexibility, the potential to allow for different interpretations of boundary objects, and (2) retaining each community's identity and practices of the social world.

Boundary objects are always used at the intersection of several ``social worlds'' or ``communities of practice''~\cite{Star1989,Bowker1999}.
Communities of practice is a concept from the world of social learning theory~\cite{Lave1991}.
They are ``groups of people who share a concern, a set of problems, or a passion about a topic, and who deepen their knowledge and expertise in this area by interacting on an ongoing basis''~\cite{Wenger2002}.
In the context of systems engineering, we can find a plethora of such communities across several dimensions (as suggested, for example, by the SAFe framework\footnote{\url{https://www.scaledagileframework.com/communities-of-practice/}}).
However, we do not necessarily focus on these official communities of practice in this paper, but also on other actors and groups in an organization.
For instance, members of a (cross-functional) team share a concern and interact on an ongoing basis as part of their daily work.

The concept of boundary objects has been analyzed in a variety of domains and with a focus on different properties~\cite{Abraham2013}.
In this paper, we focus on its use for the management of systems engineering artifacts in large-scale agile automotive contexts.
We leveraged existing studies to inform the development of our guidelines.
As Bowker and Star~\cite{Bowker1999} have described, it can be ``tempting'' to design or engineer boundary objects in an upfront manner.
This is why Levina and Vast differentiated between designated boundary objects and \emph{boundary objects-in-use}~\cite{Levina2005}.
Boundary objects are objects not only intended to be boundary objects, but actually used and managed accordingly.
They need to be both useful locally and maintain a common identity~\cite{Levina2005}.

Abraham derived a catalog of typical boundary object properties from a literature review and a focus group~\cite{Abraham2013}, and described design principles for the management of boundary objects~\cite{Abraham2017}.
As we will describe in Section~\ref{sec:RQStrategies}, we use the following principles to reason about practical tool support:
\begin{enumerate}
	\item \textbf{Interpretive flexibility}: Accept and allow for different interpretations of boundary objects. Establish boundary infrastructures to make boundary objects available~\cite{Bowker1999}, but accept that you do not need a consensus on boundary objects' meaning.
	\item \textbf{Identity preservation}: Retain each community's identity and practices of the social world. Boundary objects-in-use will be on a level that can be leveraged by different communities without restricting their identity.
	\item \textbf{Abstraction/concreteness}: Combine global, de-contextualized models with local, community-specific models. Explicate the links between models on different levels of abstraction.
	\item \textbf{Stability}: Aim for a stable and recognizable structure. Define a release management process, so that there is always one official version of the boundary object in circulation.
	\item \textbf{Modularity} based on user-based contextualization: Allow communities to use specific areas of a boundary object independently from each other.
	Provide one common view of the model for all user groups. Group information relevant to one user group. Allow users to discover information adjacent to their own area of concern.
	\item \textbf{Visualization}:
	When visualizing boundary object models, aim for semantic transparency  (so that for example the semantics of graphical symbols can be easily understood) and complexity management (so that information can be hidden to reduce the complexity).
\end{enumerate}

In our data analysis, we identified properties to boundary objects that emerged from the practical understanding of the concept.
Details about the data analysis are described in Section~\ref{sec:SystemWeaver}.
We noted that boundary object candidates are on one hand used across engineering phases, i.e., function specification, analysis, design, and verification.
For instance, software requirements can serve as such pieces of information.
We call these objects ``vertical'' boundary objects.
As a consequence of the transition to agile practices, the collaboration between these phases is strengthened.
For instance, development and verification stakeholders communicate more and are partly located in the same team.
However, the breakdown and communication of a function to several component teams is still common.
Besides the top-down collaboration, people in different teams working in the same phase need to communicate with each other.
For instance, cross-functional teams working with connected components might refer to the same design signal as their interface.
We refer to these as ``horizontal boundary objects.''

\section{Research Method} \label{sec:ResearchMethod}
Our research approach was based on design science~\cite{Hevner2004}, aiming to create guidelines as our design artifacts.
Section~\ref{sec:sec:ResearchProcess} describes the design-science research process in detail.
Section~\ref{sec:companies} presents the selected companies and participants.
Threats to validity are discussed in Section~\ref{sec:sec:Validity}.
\subsection{Design Science Research Process} \label{sec:sec:ResearchProcess}
\begin{figure}
	\centering
	\includegraphics[width=.9\linewidth]{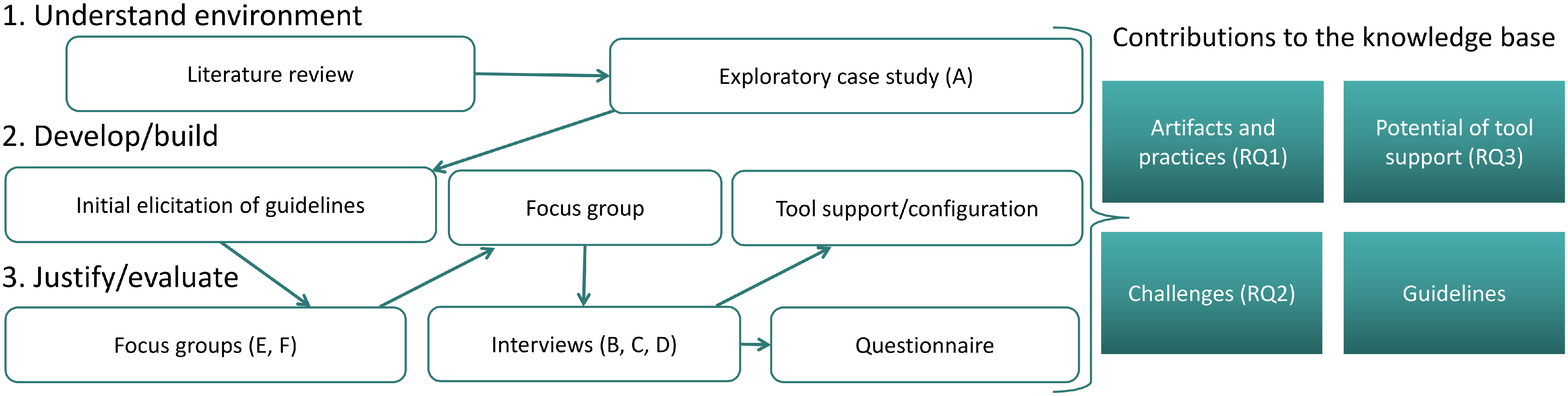}
	\caption{Overview of our design-science research process}
	\label{fig:ourmethod}
\end{figure}
Our design-science research was based on understanding the environment, developing design artifacts, and evaluating them.
The three phases with the used methods and our contributions to the knowledge base are shown in Figure~\ref{fig:ourmethod}.
In the following, we provide a detailed description of each of the phases.
\subsubsection{Understand Environment} \label{sec:sec:UnderstandEnvironment}
The initial phase was concerned with understanding the environment.
From a research perspective, we explored it with a review and discussion of related work, as presented in Section~\ref{sec:RelatedWork}.
The literature review included a systematic search for literature on the management of documentation and artifacts, especially in large-scale agile systems engineering.
Since we could not find practical guidelines on the management of artifacts in large-scale agile systems engineering, we extended the body of knowledge using a 14-week long exploratory case study.
The case study was conducted together with Company~A to get in-depth insights of artifacts, management practices, and challenges.

Our means of data collection were semi-structured interviews with participants 1--8.
The interviews' lengths were between 45 and 80 minutes, with an average of 60 minutes.
The interviewees take part in agile initiatives or were suggested by other participants because of their expertise in agile methods and systems engineering artifacts.
We created an interview guide\footnote{\url{https://www.dropbox.com/s/uprp1a4e373bdt2/interview_guide_syseng_artifacts.pdf}} including a mixture of open-ended and specific questions.

The interviews were recorded and transcribed by the main interviewer.
For the data analysis, we followed Creswell's approach for qualitative data~\cite{Creswell2008}.
For this purpose, we developed an analysis guide that also provides a detailed explanation of our analysis approach\footnote{\url{https://www.dropbox.com/s/t2gj5w4w38qx0u5/analysis_guide_syseng_artifacts.pdf}}. 
We used an \emph{editing approach} to \emph{code} the data in a systematic way~\cite{Runeson2009}.
The editing approach involves iteratively creating, revising, merging, and splitting codes.
To identify themes to report on, we discussed relations between codes and analyzed connections to related work.
\subsubsection{Develop/Build} \label{sec:sec:DevelopBuild}
The \emph{develop/build} phase was concerned with the elicitation and development of guidelines.
The initial set of guidelines was created based on our review of related work and the exploratory case study.
Additionally, we organized a focus group~\cite{Tremblay2010} that took about two hours.
In the focus group, we involved two industry representatives and two senior researchers with multiple years of experience with agile methods and documentation/artifacts.
The focus group stared with a presentation of practices, challenges, and an initial draft of guidelines.
The participants were asked to discuss the guidelines with respect to comprehensibility, consistency, and completeness.
We used several smaller iterations to refine the guidelines, discuss interdependencies, and point out relations to challenges.
The participants were also asked to discuss implications for research and practice.
We carefully took notes both on a whiteboard and on paper, asked participants to validate whether the notes adequately captured their concerns, and synthesized all notes after the focus group.
The collected information was structured as in Table~\ref{tab:Findings1}.

An additional step in the develop/build phase was to analyze the potential of tool support for the management of systems engineering artifacts in agile automotive contexts.
For this reason, we focused on a practical tool used in industrial companies, mainly in the automotive domain, analyzed its current support, and created configurations to better capture relevant aspects.
We based this phase on the data analysis and visualization of production data of Companies C and F.
Additionally, we conducted two interviews with a tool expert and meta-modeler from Company F to understand the use of the tool and boundary objects.
The results of this endeavor are presented in Section~\ref{sec:SystemWeaver}.
\subsubsection{Justify/Evaluate} \label{sec:sec:JustifyEvaluate}
The \emph{justify/evaluate} phase had the goal of validating and refining the findings.
It included two focus groups with Companies E and F that lasted about 1.5 hours.
In order to get an even broader perspective on the topic, we involved 17 testing experts in one focus group, and 25 software developers and engineers in the other one.
At the beginning of the focus groups, we introduced the topic, as well as the identified artifacts, practices, and challenges.
This information was used to start a discussion on guidelines, differences between companies, and practices when transitioning to agile methods.
Participants were encouraged to voice their concerns on all findings and notes were taken.
Directly after the focus groups, we collected all information in one document that was used to reevaluate the findings and guidelines.

An additional step in the evaluation phase were interviews with participants 9--11 that lasted for 45 minutes.
One of the interviews was conducted over the phone.
First, we presented the findings and guidelines.
Then, we discussed the findings with the interviewees, trying to find incomplete information, and focusing on implications for practice.
All interviews were conducted with at least two researchers, which enabled us to take accurate notes.
The notes were used after the interviews to phrase guidelines more clearly and get a better understanding of particular characteristics of each of the case companies.

The final step was an anonymous questionnaire.
The questionnaire\footnote{\url{https://www.dropbox.com/s/3t17e341r8mj0ch/questionnaire_syseng_artifacts.pdf}} included Likert scale questions~\cite{Likert1932} which allowed us to quantify to what extent practitioners agreed with our findings.
We presented the findings using an explanatory video of about 5 minutes in which we introduced the main terminology and concepts.
We aimed to find suitable respondents by sending the questionnaire to all participants in earlier steps, as well as additional contacts from the automotive domain.
In total, we received 41 answers and filtered out the ones that were incomplete.
In total, we analyzed 31 responses.
Tables~\ref{tab:ExperienceAgile} and~\ref{tab:AreaOfExpertise} show the respondents' experiences with agile practices and their areas of expertise.
Figure~\ref{fig:experienceyears} depicts a boxplot of the respondents' experiences with systems engineering in years.
It can be seen that the experience ranged from 0 to 30 years, with a median of 5 years.
The mean value was 9.25 years.

We analyzed the responses grouped by question, but also looked at responses per area of expertise and experience.
The open-ended answers were analyzed similarly to the analysis approach for qualitative data that we used for our interviews.
We created appropriate graphs and visualizations for many of the closed-end questions and include them in the summaries and discussions of findings for each of the research questions.

\begin{table}
	\begin{minipage}[t]{0.35\linewidth}
		\centering
		\caption{Respondents' experience with\newline agile practices, n=31}
		\label{tab:ExperienceAgile}
		\begin{tabular}{>{\raggedleft}p{.55\linewidth}>{\raggedleft}p{.2\linewidth}}
			\toprule
			Experience with agile & No.~of answers \tabularnewline \midrule
			At least a year of working with agile practices. & 17  \tabularnewline
			We have started the transition to agile but have worked with agile practices for less than a year. & 10  \tabularnewline
			None. We work in a plan-driven approach right now. & 2  \tabularnewline
			Not answered & 2  \tabularnewline
			\bottomrule
		\end{tabular}
	\end{minipage}
	\begin{minipage}[t]{0.33\linewidth}
		\centering
		\caption{Respondents' areas of expertise, n=31}
		\label{tab:AreaOfExpertise}
		\begin{tabular}{>{\raggedleft}p{.7\linewidth}>{\raggedleft}p{.2\linewidth}}
			\toprule
			Area & No.~of answers \tabularnewline \midrule
			Software Development & 6  
			\tabularnewline 
			Software/Systems Architecture & 4  
			\tabularnewline 
			Management (Project/Process) & 4  
			\tabularnewline 
			Tool Support & 3 \tabularnewline 
			Requirements Engineering & 2  
			\tabularnewline 
			Software Testing & 2  
			\tabularnewline 
			Product Owner & 2  
			\tabularnewline 
			Not answered & 2  
			\tabularnewline 
			Other (e.g., QA, tool, requirements, test, software developer and scrum master) & 6  
			\tabularnewline \bottomrule
		\end{tabular} 	
	\end{minipage}
	\hspace{0.03\linewidth}
	\begin{minipage}[t]{0.3\linewidth}
		\centering
		\caption{figure}{Respondents' experience with\newline systems engineering in years, n=31}
		\label{fig:experienceyears}
		\includegraphics[width=0.8\linewidth]{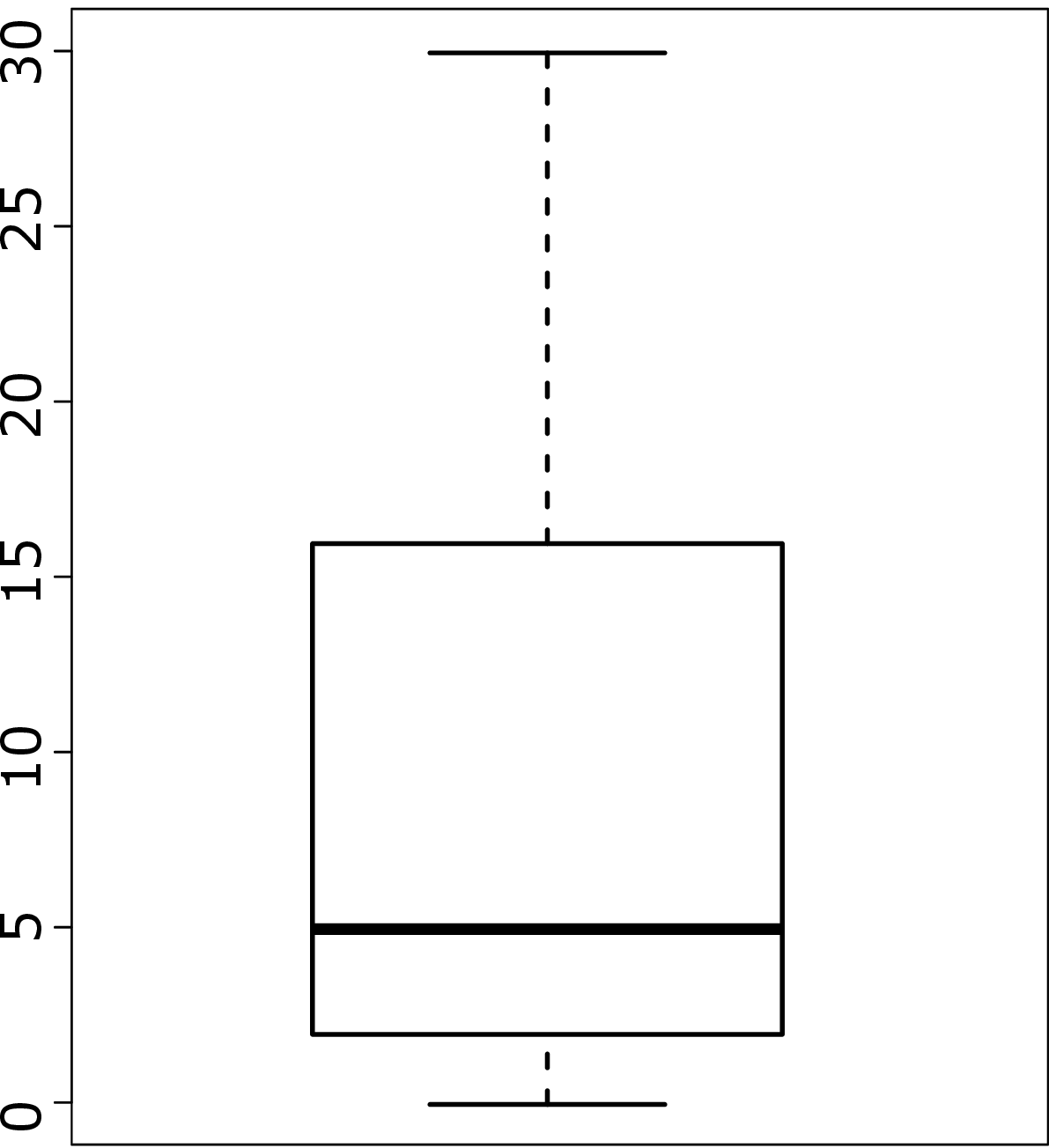}
	\end{minipage}
\end{table}

\subsection{Selected Companies and Participants} \label{sec:companies}
When selecting participants for the study, we contacted six companies to get diverse viewpoints on the topic and be able to validate our findings.
The companies were selected based on availability and existing collaborations.
In particular, the study was initiated because Company A contacted us for support with the continuous management of artifacts, which was considered a prevalent issue also in other companies.
We aimed to contact participants with different roles in the organizations and asked contact persons to reach out to and recommend other participants within their networks (snowballing)~\cite{Atkinson2001}.
In total, 53 participants were involved in interviews and focus groups.
Table~\ref{tab:Participants} gives an overview of their companies, roles, and the years of experience with systems engineering.
Additionally, 31 anonymous questionnaire respondents participated.

Our exploratory case study was conducted together with Company A, an automotive manufacturer with more than 15,000 employees.
Company A uses a mixed approach, with the top-level organization following the V-model process.
One department adapted agile practices three years before the study, whereas other departments started the transition about a year ago.
Different practices were followed in the departments.
For instance, one department used the SAFe framework~\cite{Leffingwell2007}.
Another department has been following an adapted form of Scrum~\cite{Schwaber2001} for about three years.
In Company A, several dozen cross-functional teams collaborated across at least five locations in a large-scale agile context.

Companies B, C, and E are automotive OEMs and Company F is an automotive supplier.
Company B is the only company that has finalized the transition to agile methods several years before the study.
The remaining companies have a comparable situation as Company A:
A plan-based approach is used in parts of the companies, while especially inside of development teams, agile practices have been adopted.
Scrum is the agile method most commonly followed.

Finally, Company D is a provider of an information management tool used in the automotive domain.
The company supports customers in eliciting requirements on what information to capture and how to customize the tool.
We interviewed an expert to validate our findings due to his experience that went beyond the scope of a single automotive company.

\begin{table}
	\centering
	\caption{Participants with companies, roles, and experience}
	\label{tab:Participants}
	\begin{tabular}{>{\raggedleft}p{.05\textwidth}>{\raggedright}p{.1\textwidth}>{\raggedright}p{.42\textwidth}>{\raggedright}p{.2\textwidth}}
		\toprule
		\footnotesize{No.} & \footnotesize{Company} & \footnotesize{Role} & \footnotesize{Experience} \tabularnewline \midrule
		1 & A & \ndxR & $>$ 20 years \tabularnewline
		2 & A & \txR & $>$ 17 years \tabularnewline
		3 & A & \jxR & $>$ 15 years  \tabularnewline
		4 & A & \vxR & 19 years \tabularnewline
		5 & A & \ruR &  $>$ 25 years\tabularnewline
		6 & A & Team leader (functional dev.) & 17 years \tabularnewline
		7 & A & \muxR & 15 years \tabularnewline
		8 & A & \raR &  $>$ 2 years \tabularnewline \midrule
		9 & B & Software architect &  9 years \tabularnewline		
		10 & C & Systems architect & $\sim$ 20 years \tabularnewline
		11 & D & Chief Technical Officer &  $>$ 30 years \tabularnewline \midrule
		12--28 & E \& F & Tester & 1--30 years \tabularnewline
		29--53 & E \& F & Software developer/engineer & 1--30 years \tabularnewline 																
		\bottomrule
	\end{tabular}
\end{table}

\subsection{Threats to Validity} \label{sec:sec:Validity}
This section discusses Maxwell's validity threats for qualitative research~\cite{maxwell2012qualitative}:
\paragraph*{ \em Descriptive Validity:} \label{sec:sec:Descriptive}
Descriptive validity is concerned with the extent to which the analyzed data is accurate.
Our analysis process is based on transcripts from interviews and focus groups.
Moreover, our questionnaire responses include both quantitative and qualitative data (from closed and free-text questions).
It might have influenced the meaning that transcripts do not allow the analysis of the tone of voices, irony, or the speed of speech.
To mitigate potential biases because of this, we also included timestamps in the transcriptions.
Timestamps made it possible to revisit parts of the recordings in case we wanted to clarify the meaning of unclear statements.
We also contacted all participants from earlier steps for the validation of our findings in a questionnaire.

\paragraph*{ \em Interpretative Validity:} \label{sec:sec:Interpretative}
Interpretative validity is concerned with data being biased by our own or by our participants' interpretation based on the use of language or underlying concepts.
An attempt to mitigate biases to interpretative validity was to establish a common terminology, adhere to the participants' terms as much as possible, and presenting the context and concepts of the study to involved practitioners.
Another mechanism was paraphrasing of participants' statements which helped to clarify unclear statements.
A potential threat to interpretative validity was that we had to translate parts of the interviews from Swedish to English which might have come with translation issues.
However, conducting interviews in the participants' mother tongue allows them to express themselves in a more differentiated way and better describe their perspectives.

\paragraph*{ \em Theoretical Validity:} \label{sec:sec:Theoretical}
This form of validity is concerned with underlying theories influencing the research conducted.
One strategy of mitigating this issue was to involve other researchers in discussions of our findings.
We also aimed to mitigate bias and preconceptions by using a thorough evaluation of our findings using focus groups, interviews, and a questionnaire (see Section~\ref{sec:sec:JustifyEvaluate}).

\paragraph*{ \em Researcher Bias:} \label{sec:sec:ResearcherBias}
Threats to researcher bias exist in any qualitative study since all conclusions are influenced by the researchers' preconceptions and background.
We aimed to reduce researcher biases by critically reflecting on them, especially during the analysis of our findings.
A mitigation strategy was the thorough evaluation using multiple methods.
It also helped to include a variety of practitioners with different backgrounds in our participant selection, which allowed us to analyze different perspectives on the topic.

\paragraph*{ \em Reactivity:} \label{sec:sec:Reactivity}
Reactivity is concerned with the researchers' influence on participants and the setting of the study.
It is a threat to be considered in any study, as there are many influencing factors (e.g., related to how questions are formulated or intoned).
Our interview guide was reviewed by fellow researchers to identify and rephrase potentially ambiguous questions.

\paragraph*{ \em Generalizability:} \label{sec:sec:Generalizability}
Generalizability is concerned with the transferability of our results to other cases.
Our study was conducted in collaboration with six companies and more than 50 practitioners.
In different parts of the data collection, different participants were involved.
We carefully validated our guidelines and findings in multiple steps with different automotive companies to analyze their generalizability.
We expect our findings and guidelines to be generalizable to other domains, especially in large-scale agile contexts.
However, dedicated studies are required for this purpose, as will be discussed in Section~\ref{sec:Discussion}.

\section{Findings} \label{sec:Finding}
Our study focuses on understanding practices and challenges with the continuous management of systems engineering artifacts, facilitating the management by designing guidelines, and describing how a practical tool can be leveraged to support the guidelines.
This section presents our findings related to artifacts and practices (Section~\ref{sec:RQArtifacts}), challenges (Section~\ref{sec:RQChallenges}), and designed guidelines for practitioners (Section~\ref{sec:RQStrategies}).
At the end of each subsection, we summarize and discuss the findings with related work.
In Table~\ref{tab:Findings1}, an overview of our findings, their implications, and discussed related work is provided.
\subsection{Artifacts and Practices (RQ1)} \label{sec:RQArtifacts}
\begin{table*}
	\centering
	\caption{Categorization of identified artifact types with current practices, management effort, relevance, and affected organizational areas}
	\label{tab:CategorizationOfArtifacts}
	\begin{supertabular}{>{}p{.03\textwidth}>{\raggedright}p{.11\textwidth}>{\raggedright}p{.19\textwidth}>{\raggedright}p{.22\textwidth}>{\raggedright}p{.15\textwidth}>{\raggedright}p{.16\textwidth}}
		\toprule
		Scope & {Artifact Type} & {Current practices} & {Effort for stakeholders} & Relevance & Main Area(s) \tabularnewline
		\midrule
		\multirow{3}{*}{{\rotatebox[origin=r]{90}{Boundary object candidates}}} &Architecture Models \& Descriptions & Practices differ in departments.
		Not consistently managed as a living document. & Handle scattered information. Counteract architecture degradation. & ``Big picture'' of the system's structure and design decisions. & Architecture, cross-functional teams \tabularnewline \cmidrule{2-6}
		& High-level Requirements & Used in many parts of the organization.
		To some extent replaced by user stories. & Find an appropriate level of detail so that information is useful for stakeholders. & High-level functionality and purpose, common reference of product. & Product management, architecture, cross-functional teams \tabularnewline \cmidrule{2-6}
		& Variability Information & Spread throughout several tools and systems.
		& Manage complexity and deal with scattered information. & Central for all phases of the lifecycle. & All areas \tabularnewline \midrule[0.145em]
		\multirow{3}{*}{\rotatebox[origin=r]{90}{Locally relevant artifacts}} & Documen-\\tation & Varying, depending on processes. Mostly text documents. & Create documents on an appropriate level of detail, deal with changes, motivate developers. & Compliance with legislation and standards. & Cross-functional team \tabularnewline \cmidrule{2-6}
		& Low-level Require-\\ments \& Tests & Less detailed descriptions, more test cases than during plan-driven development. & Keep requirements up-to-date. Find the right level of detail to avoid overspecification. & Low-level quality assurance. Comm.~with suppliers. & Cross-functional team \tabularnewline \cmidrule{2-6}
		& Simulink models \& signal databases & Used in teams to generate code and signal databases. & Ensure that people are aware of (otherwise implicit) rationales and intentions of models. & Code generation on a low level, ensured consistency. & Cross-functional team \tabularnewline
		\bottomrule
	\end{supertabular}
\end{table*}

This section presents our findings related to RQ1: \rqArtifacts

Table~\ref{tab:CategorizationOfArtifacts} shows a summary of artifacts, management practices, the main effort of managing them, artifacts' relevance, and the affected organizational areas.
It should be noted that we discussed general artifacts and types, rather than concrete artifact instances.
In some cases, interviewees also mentioned concrete documents to support their responses with examples.

In the discussion of this subsection, we elaborate on the leftmost column in Table~\ref{tab:CategorizationOfArtifacts} ``Scope'': We found that some artifacts can be seen as boundary objects, which has a high influence on how they are managed and used.
\subsubsection{Architecture Models and Descriptions}
In the discussion of artifacts in large-scale agile automotive contexts, architecture models and descriptions stood out because of their potential to convey a ``big picture'' of the system and its design decisions:

\begin{aquote}{\ndxR}
	If someone wants to get the full picture, then it's impossible to not have some kind of understanding how things are working, how things are coupled and connected. [...]
	And that's why the architecture is so important [...], to communicate about the product.	
\end{aquote}

Architecture models are used by several actors in the organization, most prominently by developers and architects.
The tools in which architecture information is stored vary: In some departments, a model-based systems engineering tool is used, while text documents are common in other departments.
So far, Company A has faced difficulties with managing a model of the high-level architecture as a living document.
These struggles are further impeded by heterogeneous information management approaches in distributed departments.
Architecture degradation is one of the consequences that practitioners strife to counteract.
\subsubsection{High-level Requirements}
Requirements are artifacts mentioned by all interviewees, which underlines their importance in the automotive domain.
In parts of Companies A and C, plan-driven practices are still used, including formal processes to hand over requirements specifications.
In other parts, requirements specifications are replaced by user stories.

\Tx stressed that \emph{system requirements} on a high level are crucial to specify the functionality and interaction of functional components.
This high-level information is used as a common reference for product management, development, and testing.
\Vx described the biggest challenges as finding the ``right level of detail'' and communicating this information to involved stakeholders.

\subsubsection{Variability Information}
Variability is a big challenge in automotive and this complexity needs to be managed throughout the systems' lifecycles.
\Ndx mentioned that he used the systems engineering tool most commonly to find variability information.
However, besides the systems engineering tool, also other tools and databases are used to store variability information.
The scattered information comes with a higher effort to identify and collect relevant information.
For this purpose, one of the departments' visions states that main features and variants should be documented consistently ``as a reference for development, production, and maintenance.''

\subsubsection{Documentation}
Authorities, standards, and customers require certain documentation.
This documentation is still required when transitioning to agile.
According to \tx, documentation describing how customers can configure the vehicle's functions needs to be created.
This documentation is supposed to be written by developers who rather invest time in actual development.
Similar issues are faced when having to create documentation for authorities or compliance purposes.
In most organizational groups, this documentation is written in textual form by developers or engineers.
Besides motivating authors to compose this documentation, it is an effort to find an adequate level of detail and deal with change.

\subsubsection{Low-level Requirements and Tests} \label{sec:sec:LowLevelReq}
In plan-driven approaches, a high level of detail was used to describe the system's logical design in a systems engineering tool.
Taking a requirement as an example, \ndx concluded that ``if you put a semicolon in the end, you could almost generate code from it.''
Low-level requirements and test cases are commonly used when suppliers are contracted.
Also for in-house development, these artifacts can be leveraged for low-level quality assurance.
It should be noted that detailed specification also comes with a higher maintenance effort to keep these artifacts up to date and in sync with the code.
\Ndx saw the high level of detail as problematic as some information is ``so close to the software code level that [he is] questioning'' its relevance.
\Tx reported that it is difficult to convince teams to document low-level requirements and keep them in sync with test cases and the code.
However, while the chief architect observed that requirements were neglected when transitioning to agile, the importance of test cases increased.
The amount of test cases has increased and to some extent, ``test cases themselves start to be the requirements.''
However, the chief architect stressed that this only works because there has not been any major development of new functions.

Note that analyzing artifacts from an inter-organizational angle, low-level requirements can also be seen as boundary objects between an OEM and suppliers.
Our perspective are different social worlds in an organization, which is why we regard low-level requirements and test cases as locally relevant artifacts.
\subsubsection{Simulink and Signal Database Models}
When discussing relevant artifacts, our interviewees mentioned Simulink models, as well as models used to generate signal databases for communication networks.
\Tx stated that developers are motivated to use Simulink models for code generation, since they are directly in relation to the generated executable code.
On the contrary, low-level requirements need to be kept in sync manually and inconsistencies do not directly lead to issues in the code execution.
\Ndx reported that ``it must be updated. Otherwise, it will not work.''

The discussed models belong to the group of prescriptive artifacts, i.e., artifacts used to generate code or other artifacts, mostly by developers and engineers.
It is easy to see their usefulness for subsequent phases and the relation to the final system.
However, \ra stressed that, for prescriptive artifacts, ``you capture how it's done but not why.''
For this reason, \ao pointed out that prescriptive models should be complemented with high-level descriptions:
\begin{aquote}{\aoR}
	The model gives you the solution. This is how to do it. And then you don't know if that is the actual requirement. So there needs to be, I think, some written text on a higher level [...], saying what the intention is.
\end{aquote}
Our interviewees saw it as critical to set models into a context and clarify their intentions and rationales.
\subsubsection{Summary and discussion of findings} \label{sec:rq-artifacts:discussion}
\begin{figure}
	\centering
	\includegraphics[width=0.8\linewidth]{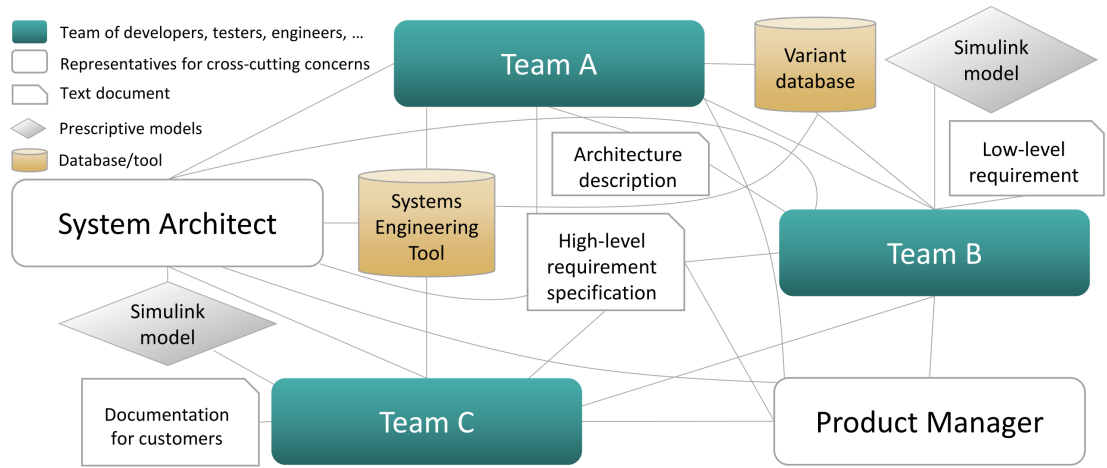}
	\caption{Exemplary actors and their relations in systems engineering}
	\label{fig:socialworlds}
\end{figure}

In this section, we have presented architecture models, high-level requirements, and variability information and how they are managed.
These artifacts are used and managed in different organizational groups to create a big picture of the system.
However, we also found artifacts that are used within a cross-functional team (i.e., documentation, low-level requirements and tests, Simulink models, and signal databases).
We summarize these findings as follows:
\begin{finding}\label{fin:F1}
	Relevant artifacts are either important to support system-level thinking in an organization or play a role inside a team.
\end{finding}

We see architecture models, high-level requirements, and variability information as boundary object candidates.
In this paper, artifacts used inside a team are referred to as \emph{locally relevant artifacts}.
Our findings are in line with related work.
For instance, architecture models and descriptions have been described to be important boundary objects before~\cite{Pareto2010,Abraham2017}.
Especially in the transition to large-scale automotive agile practices, the need to explicitly document the architecture has been proclaimed~\cite{Pretschner2007, Prause2012}.
In the questionnaire, 17 respondents indicated in the free-text field that they see requirements as boundary objects, and 6 mentioned architecture models.
Interface descriptions were mentioned by 6 participants, and signals specifications/lists and communication databases were mentioned by 5 of our 31 participants.
For locally relevant artifacts, low-level requirements were named by 9 respondents, Simulink models by 7, and test cases/reports/specifications by 8 respondents.
Other mentioned artifacts were user stories, the product backlog, done criteria, or code.

Note that many of the locally relevant artifacts are of prescriptive nature, describing a subject that does not exist yet.
They are typically used to generate code or other artifacts.
In our study, Simulink models, signal databases, and test cases were mentioned.
With respect to the management of information, we found a higher willingness to keep prescriptive artifacts updated as they are directly related to other artifacts.
\begin{finding}\label{fin:F2}
	Stakeholders' motivation to manage prescriptive models is higher than dealing with descriptive artifacts.
\end{finding}
Selic confirms modeling languages that produce executable models as a promising enabler for agile documentation~\cite{Selic2009}.
In the automotive domain, the quality of prescriptive models is high, as developers see the direct impact on the implementation and keep them up to date~\cite{Eliasson2015}.

\begin{figure}
	\centering
	\includegraphics[width=0.9\linewidth]{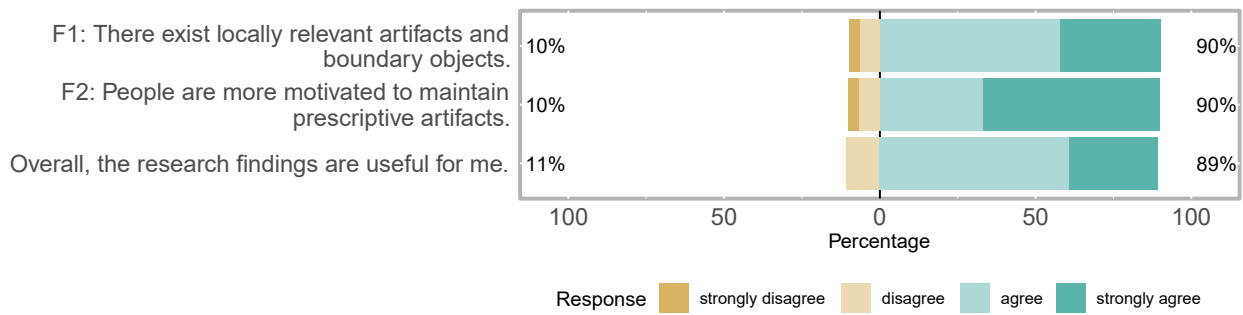}
	\caption{Questionnaire responses regarding F1 and F2, n = 31}
	\label{fig:findings}
\end{figure}

Figure~\ref{fig:findings} depicts the questionnaire responses regarding F1, F2, and the usefulness of our findings.
A majority of our 41 anonymous respondents agrees with the findings and considers them useful.
Of the participants who have worked with agile approaches for at least a year, 94.12\% agree or strongly disagree with the fact that there exist locally relevant artifacts and boundary objects (F1).
Moreover, all participants currently working with plan-driven methods regarded the findings as useful.
We could not find any differences with respect to various roles giving different responses to these questions.
Also in the validation focus groups, we identified the classification to be of relevance for other automotive companies.

The presented information can be understood in the context of the theoretical foundation presented in Section~\ref{sec:Theoretical}.
To illustrate the applicability of actor-network theory in our context, Figure~\ref{fig:socialworlds} shows an example of human and non-human actors and their relations in systems engineering.
It represents the organization of Company A, but abstracts from details and only depicts a small subset of the actors and relations.
Actors are shown in boxes that are connected using lines.
Agile teams comprise several responsibilities: Development, testing, and engineering related to a specific feature.
Besides, there are actors representing cross-cutting concerns, e.g., systems architecture or product management.
Other actors are databases or tools, as well as text documents or models.
The lines between actors represent relations between them.
A systems engineering tool is used by parts of the company, storing architecture models, variability information, and requirements on several levels.
Moreover, high-level requirement specifications and architecture descriptions are shared between several actors.
Furthermore, there are pieces of information used within a smaller group of actors (e.g., documentation for customers or low-level requirements).
It should be noted that a lot of actors and relationship are not shown in this exemplary figure.
For instance, there might be groups of members of feature teams forming an official or unofficial community of practice (e.g., related to testing practices) across the depicted team borders.
Moreover, the relationships between actors are not homogeneous, but characterized by different power relations, dependencies, and potentially conflicting interests.

Different actors in the organization need to align their interests to collaborate successfully.
As we described in Section~\ref{sec:Theoretical}, translation in the original sense~\cite{Callon1986} is initiated by one actor and aims to establish the other actors as ``allies''.
In a hypothetical scenario, a product manager specifies a new requirement for Team B whose implementation also affects artifacts created by Team C---and wants it to be implemented immediately.
A system architect sees the need to redesign the architecture as the proposed requirement would go with a performance decrease and the necessity to reconsider architecture decisions.
Feature Teams B and C do not understand in detail how the new requirement would affect each other's parts and neither of the teams feels responsible for the connecting interfaces.
A discussion starts and the product manager aims to create allies in the organization by stressing the customer needs fulfilled by the new requirement.
After a successful translation, the other actors support the product manager's interest and the product manager establishes him-/herself as a spokesman related to this concern.
At the same time, the system architect attempts to conduct a translation for architecture redesign and convincing the feature teams to adhere to the new architecture.
In practice, the attempted translations are partly supported by power relations in the (management) hierarchy.
The theoretical context teaches us that knowledge translations that occur in parallel can be facilitated by boundary objects, as we described in Section~\ref{sec:Theoretical:ActorNetwork}~\cite{Abraham2017,Carlile2002}.
These boundary objects do not only impact a single team, but rather create a common identity supporting translation, while allowing for interpretative flexibility~\cite{Star1989}.

\subsection{Challenges With Managing Artifacts (RQ2)} \label{sec:RQChallenges}
The management of artifacts is related to practical difficulties that impede successful knowledge translation among actors.
In this section, we elucidate these difficulties and give answers to RQ2: \rqChallenges

\subsubsection*{Ch1: Diversity vs alignment} \challenge\label{cha:ch1}
Having to balance the trade-off between autonomy and alignment of cross-functional teams was reported as challenging:

\begin{aquote}{\jxR}
	You want that the teams work similarly when it comes to questions about the whole system.
	But at the same time, we want that the teams are as autonomous as possible so that they work at their own pace.
\end{aquote}

This trade-off impacts also how artifacts are used and managed.
Several of the managers we interviewed stated that they aim to allow for diversity between teams, but are also required to find appropriate points to align their work and facilitate the integration of the final product.

\subsubsection*{Ch2: Degradation of artifacts} \challenge\label{cha:ch2}
When moving to agile practices, the issue of artifact degradation has been observed.
\Tx explained this challenge as follows:
\begin{aquote}{\txR}
	We thought the teams would try to influence the complete product. But [they] have been focused inwards, on their deliverables. When they looked outside then to see where their functionality fits. There is no drive from the organization itself to change the architecture.
\end{aquote}

This challenge is partly due to the fact that ownership and responsibilities for artifacts are unclear and therefore not naturally assumed.
According to \ra, the issue of degradation can be mainly observed in architecture models and requirements.

\subsubsection*{Ch3: Mix of plan-driven and agile methods} \challenge\label{cha:ch3}
In all of the participating companies, parts of the organization transition to agile methods at different points in time.
When collaborating with non-software teams, it is more difficult to adopt agile practices.
\Jx stressed the importance of mechanics and the synchronization with their development cycles.
Software, electronics, and mechanics cannot be integrated on a daily basis but require appropriate representations, e.g., in models.
Practitioners need to find agile ways of working within one team or department and synchronization mechanisms with external organizational units.
\subsubsection*{Ch4: Deciding on important artifacts} \challenge\label{cha:ch4}
Managing artifacts is an effort, which is why one needs to identify what artifacts are crucial to be created and managed.
The challenge of deciding on important artifacts was stressed by \ao and \vx:
\begin{aquote}{\vxR}
	A lot of things you do in an iteration is only needed to communicate between members of a team. So you have to think carefully about what is most important to maintain the software and correct it.
\end{aquote}
\IntC faced similar problems in his company:
Besides identifying what artifacts are relevant, also the level of granularity is difficult to determine.
Currently, architectural models are on a too detailed level which is planned to be reduced in the future.

\subsubsection*{Ch5: High staff turnover} \challenge\label{cha:ch5}
The quality of systems engineering artifacts depends a lot on the individuals in charge.
\Jx stated that a high staff turnover is a problem when documenting the architecture:
\begin{aquote}{\jxR}
	We have had a high turnover of consultants who write this.
	So that functions that we created were changed because the new ones did not understand them.
\end{aquote}
The staff turnover is one of the motivators to reconsider approaches to knowledge management.
\subsubsection*{Ch6: Different locations and backgrounds} \challenge\label{cha:ch6}
Agile methods traditionally rely on face-to-face communication.
\Ao pointed out that following an agile method ``requires that you can talk just like this.''
Different time zones complicate the scheduling of meetings.

Also heterogeneous backgrounds are an issue, as different terms are used and concepts are understood.
\Jx reported the challenge that the plan-driven top-level organization does not have ``a full understanding of software'', due to different professional backgrounds.
\Ru stated that it can be a challenge to ``fight the control culture'' imposed by management.
This interviewee underlined the necessity of actively investing in maintaining an agile culture.

\subsubsection{Summary and discussion of findings} \label{sec:rq-challenges:discussion}
\begin{figure}
	\centering
	\includegraphics[width=0.85\linewidth]{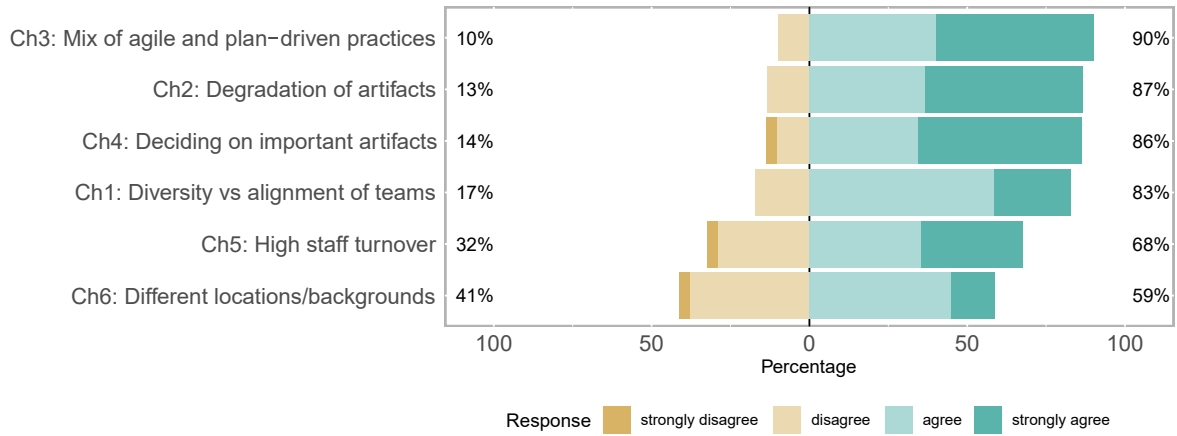}
	\caption{Questionnaire responses regarding the challenges, n = 31}
	\label{fig:challenges}
\end{figure}

Several challenges were identified when analyzing the management of artifacts in agile automotive contexts:
Aligning diverse teams, degradation of artifacts, mixing plan-driven and agile methods, identifying important artifacts, staff turnover, and different locations and backgrounds.
Figure~\ref{fig:challenges} shows the questionnaire responses regarding to what degree participants agree with the identified challenges.
Many participants have confirmed the challenges, not only questionnaire respondents, but also interviewees and focus group participants.
For instance, a tester from Company E stated that they ``have seen virtually all challenges in [their company].''
Differences can be seen depending on the experiences of the respondents:
All of the respondents stating that they disagree with Ch1 (diversity vs alignment) or Ch4 (deciding on important artifacts) had less than 4 years of experience with systems engineering or had been working with agile methods for less than a year.
Ch6 (different locations/background) was not regarded challenging by 41\% and was reported as a challenge by 59\%.
For this challenge, a difference can be seen with respect to the roles of the participants:
Of the participants working with software/systems architecture, 100\% agreed or strongly agree with the challenge.
On the other hand, 83.3\% of the respondents working with software development disagreed or strongly disagreed with the statement that they faced Ch6.
Also all product owners report that they do not consider Ch6 challenging.
These results indicate that the roles and focus of individuals impact how challenges are perceived.
Our analysis of challenges confirms several of the issues reported by Petersen and Wohlin~\cite{Petersen2010}, e.g., the management overhead to align the work of different teams.
They stated that the move to agile reduced the need for documentation, but did not point out any challenges related to the management of artifacts.
However, the challenges of deciding on important artifacts and handling artifact degradation were confirmed by other related studies~\cite{Lagerberg2013,Ruping2003,Hanssen2009}.
Moreover, it is indeed common that a mix of methods is used and not the entire organization moves to agile practices at once~\cite{Kuhrmann2018}.

\begin{table} [b]
	\centering
	\caption{Mapping of challenges to boundary object properties (Section~\ref{sec:Theoretical:Properties})}
	\label{tab:challenges}
	\begin{tabular}{@{}>{\raggedright\arraybackslash}p{.15\textwidth}>{\centering\arraybackslash}p{.09\textwidth}>{\centering\arraybackslash}p{.09\textwidth}>{\centering\arraybackslash}p{.09\textwidth}>{\centering\arraybackslash}p{.12\textwidth}>{\centering\arraybackslash}p{.09\textwidth}@{}>{\centering\arraybackslash}p{.12\textwidth}@{}}
		\toprule
		& Ch1 (diversity vs alignment) & Ch2\newline (artifact degradation) & Ch3 (mix of plan-driven and agile) & Ch4 (deciding on important artifacts) & Ch5 \newline (high staff turnover) & Ch6 (different locations \& backgrounds) \tabularnewline \midrule
		Interpretive flexibility & X &  &  &  &  & X \tabularnewline \midrule
		Identity preservation & X &  &  &  &  & X \tabularnewline \midrule
		Abstraction/concreteness &  & X &  & X &  &  \tabularnewline \midrule
		Stability &  &  & X &  & X & X  \tabularnewline \midrule
		Modularity & X & X &  & X &  &  \tabularnewline \midrule
		Visualization &  &  &  &  & X &  \tabularnewline \bottomrule
	\end{tabular} 
\end{table}
The challenges relate to boundary object properties that we described in Section~\ref{sec:Theoretical:Properties}.
Table~\ref{tab:challenges} shows an overview of these relations that we will describe in the following.
{Challenge 1 (diversity vs alignment) can be understood in the context of interpretive flexibility and identity preservation.
	The meaning of boundary objects should not be dictated, but allow for different interpretations.
	The identity of each social group needs to be preserved to allow for diversity.
	At the same time, boundary objects provide a way for different actors to collaborate without consensus.
	In this context, modularity helps end users to reuse information as needed, while allowing for flexibility in their own module.}
{Challenge 2 (artifact degradation) relates to the abstraction and modularity properties of boundary objects.
	Teams should be supported by understanding different levels of abstractions, i.e., both the local and global picture.
	The focus should not only lie ``inwards'', but also the system-level understanding should be supported to see how different modules are related and can be reused.}
Challenge 3 (mix of plan-driven and agile methods) can be supported by having stable boundary objects that can work as a synchronization mechanism to external units.
This way, released and versioned boundary objects can be a ``contract'' between groups working with different methodologies.
Challenge 4 (deciding on important artifacts) relates to the property of modularity: Boundary objects should be accessible in a way appropriate for the user-based context.
Understanding the modularization of a system helps to identify what artifacts are important to maintain.
Moreover, the property of abstraction/concreteness helps to define boundary objects on a suitable level of detail.
Challenge 5 (high staff turnover) can be mitigated by having stable boundary objects whose changes are managed and recognizable over time, also by different stakeholders.
Mechanisms for visualization can help new stakeholders to familiarize faster in the process of understanding the existing information.
Challenge 6 (different locations and backgrounds) can be mitigated by having a stable and recognizable structure for different social groups (stability).
This challenge relates to Ch1, as it also is concerned with aligning diverse groups in a suitable way.
For this reason, Ch6 also connects to the importance of having interpretive flexibility and preserving the identity of different groups in distributed locations.

\subsection{Guidelines to Manage Artifacts} \label{sec:RQStrategies}
\begin{figure}[b]
	\centering
	\includegraphics[width=0.85\linewidth]{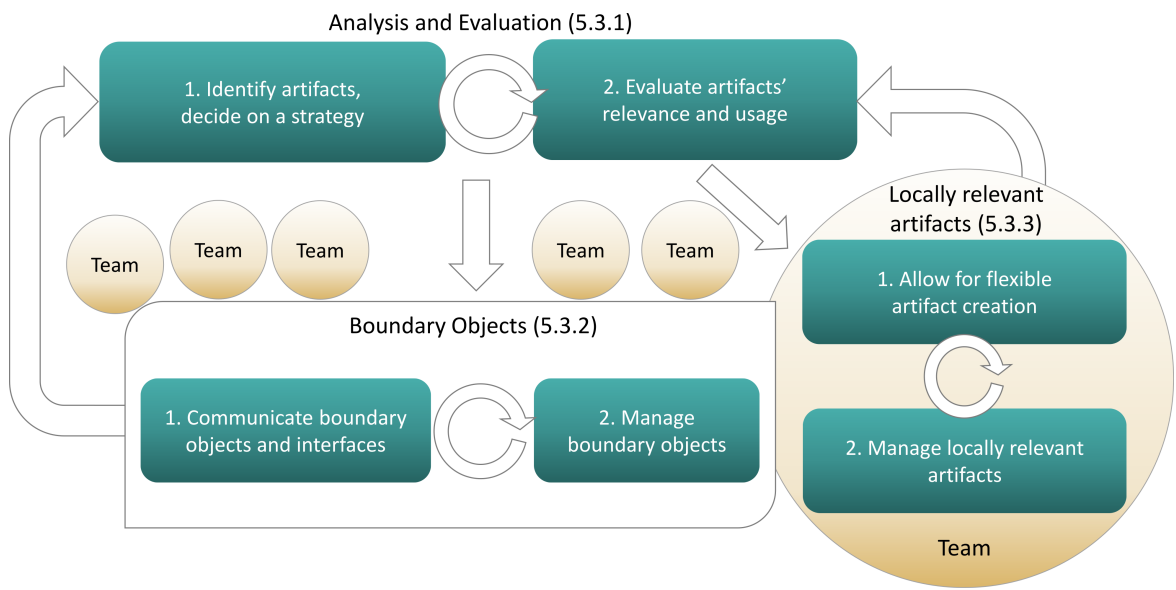}
	\caption{Guidelines to manage artifacts}
	\label{fig:strategies}
\end{figure}
Based on our understanding of artifacts, practices, and challenges, we developed guidelines to support the management of systems engineering artifacts in agile automotive contexts.
The guidelines are derived from our empirical data, as well as from the theoretical foundation of the area as described in Section~\ref{sec:Theoretical}.
Figure~\ref{fig:strategies} gives an overview of our guidelines that are depicted in boxes.
Arrows show transitions between guidelines and loops show that they should be executed iteratively.
We describe guidelines grouped by three areas, i.e., analysis and evaluation (Section~\ref{sec:sec:AnalysisEvaluation}), boundary objects (Section~\ref{sec:sec:BoundaryObjects}), and locally relevant artifacts (Section~\ref{sec:sec:LocalObjects}).
\subsubsection{Analysis and Evaluation} \label{sec:sec:AnalysisEvaluation}
We found that relevant artifacts need to be periodically identified and reevaluated, as the goal is to leverage boundary objects-in-use that are beneficial to the involved stakeholders~\cite{Levina2005}.

\vspace{0.5em}
\textbf{Identifying boundaries and deciding on a strategy} is an important first step to understand how artifacts should be managed.
\Vx and \ru stress the importance to understand who uses what information for what reason and with what challenges.
Also the levels of abstraction should be clarified and it is recommendable to explicate the links between models on different levels of abstractions~\cite{Abraham2017}.
Table~\ref{tab:CategorizationOfArtifacts} can be used as a template to categorize artifacts.
While we approached the topic from a large-scale agile lens, it should ideally cover other parts of the organization and lifecycle (e.g., maintenance, sales, ...).

Based on these findings, we phrase the first guideline:
\begin{guideline}\label{gui:define_strategy}
	Involve stakeholders from different areas to identify what artifacts are boundary objects and locally relevant artifacts. Find out how and by whom they are used, who should be responsible at what point in time, and how traceability can be established. A structure like Table~\ref{tab:CategorizationOfArtifacts} can support this task.
\end{guideline}
The following aspects should be clarified in this step:

\paragraph*{ \em Relevance and intentions of capturing artifacts:}
To ensure that artifacts are actually in use, the intentions of information need to be clarified, as \vx explained.
An agreement should be found as to what level of detail is needed to avoid ``over-specification.''
In Table~\ref{tab:CategorizationOfArtifacts}, we capture this aspect in the ``Relevance'' column.

\paragraph*{ \em Affected areas and stakeholders of artifacts:}
\Vx and \jx reported the need to capture responsibilities and ownership.
\Tx gave several examples of teams that varied the level of internal documentation over time, for instance, due to the varying complexity of the developed functionality and the team members' backgrounds.
In Table~\ref{tab:CategorizationOfArtifacts}, we show affected organizational areas in the ``Area'' column.

\paragraph*{ \em Strategies for traceability management:}
Traceability between artifacts should be part of the strategy, especially to specify how models on different levels of abstraction are related~\cite{Abraham2017}.
Six interviewees mentioned the importance of traceability.
For instance, \ndx stressed the importance of traceability for change impact analysis.

\vspace{0.5em}
\textbf{Evaluating artifacts' relevance and usage} at periodic intervals is needed as classifications of information develop over time (see Section~\ref{sec:Theoretical:ResidualCategories}).
Although boundary objects typically possess a stable and recognizable structure~\cite{Abraham2017}, it can happen that boundary objects lose their common identity~\cite{Levina2005}.
\Ra mentioned that features to automatically analyze the usage of information might help with these activities.

\begin{guideline}\label{gui:evaluate_strategy}
	Make sure that you evaluate artifacts' relevance and usage at frequent intervals. Depending on how the system evolves throughout its lifecycle, artifacts might become more relevant or should be discarded. Boundary objects could be converted into locally relevant artifacts and vice versa.
\end{guideline}

\subsubsection{Management of Boundary Objects} \label{sec:sec:BoundaryObjects}
Boundary objects should support the creation of a common identity, while allowing for flexible interpretations (see Section~\ref{sec:Theoretical:Properties}).
Therefore, crucial steps are to (1) communicate boundary objects and interfaces, and (2) to manage boundary objects.

\vspace{0.5em}
\textbf{Communicating boundary objects and interfaces} is relevant for actors to be conscious about the interrelations of artifacts and to create a shared understanding.
Communication is partly supported by a systems engineering tool that supports traceability, change management, and views for different groups.
In this way, practitioners can specify artifacts and create explicit links to objects in other contexts.
For instance, there exists a top-level object representing the high-level architecture, including software and hardware components and their connections.
Following Abraham's design principle of abstract/concreteness (see Section~\ref{sec:Theoretical:Properties}), the abstract top-level object can then be linked to more concrete models.
\Vx stressed that the classification of information and abstraction levels needs to be explicit:
One should ``define why things are connected. Otherwise, you start to add [information] but you actually break an intention.''

Decisions related to boundary objects can be taken on two levels, supporting different levels of abstraction.
\Mux explained that in a top-level community, decisions concerning boundary objects could be discussed, and then propagated to lower level groups ``to spread the knowledge.''
This is partly to ensure that the structure is kept stable and recognizable~\cite{Abraham2017}.
To mitigate issues of scattered information and benefit from visualization features, a common tool is recommendable.
User-based contextualization can help teams see their parts in the ``big picture'' and satisfy different information needs, as was also confirmed by the CTO from Company D, different views of the data.

We phrase the following guideline:
\begin{guideline}\label{gui:manage_bos}
	For each boundary object within your organization, establish a group of representatives from affected teams to discuss issues and later propagate that information. Store information in an accessible tool.
\end{guideline}

\vspace{0.5em}
\textbf{Managing boundary objects} and refining them over time is important.
\Vx and Levina and Vast's research~\cite{Levina2005} pointed out the importance of having boundary objects-in-use.
It is important to capture the right information at the right point in time, as we express in Guideline~\ref{gui:manage_architecture}:
\begin{guideline}\label{gui:manage_architecture}
	Find a lightweight and flexible approach to defining high-level artifacts upfront. Exploit this information to make impact analysis and changes, to avoid suboptimal decisions. With time, the artifacts should be refined and become more mature.
\end{guideline}

Ideally, boundary objects should have a stable and recognizable structure, as we explained in Section~\ref{sec:Theoretical:Properties}.
A change management and release management process can help to ensure control over boundary objects.

\subsubsection{Management of Locally Relevant Artifacts} \label{sec:sec:LocalObjects}
Locally relevant artifacts are managed within a well-connected team of actors.
We found that practitioners should
(1) allow for flexible ways of artifact creation, and (2) manage locally relevant artifacts.
\vspace{0.5em}

\textbf{Allowing for flexible artifact creation} is necessary when transitioning to agile practices: The order of artifact creation can change and become more flexible.
For instance, low-level requirements might not have to be specified before starting the implementation and development of test cases, as \jx stated.
At a later point in time, requirements are needed for maintenance, compliance with safety regulations, etc.
Our interviewees reported that artifacts are typically not needed at the beginning of the development, but at a later point in time:
\begin{aquote}{\raR}
	[There is] the concept phase, where it is very fluent and agile. Then settling it down into the formal versioning. And then sometime it [...] lies around, but is still needed.
\end{aquote}

We found that especially documentation or low-level requirements and tests are not required in the concept phase, but need to be in place at a later point in time for maintenance and aftermarket.
As developers are more motivated to maintain prescriptive artifacts, these are preferable whenever possible.
\begin{guideline}\label{gui:creation_locally_rel}
	Produce locally relevant artifacts, especially those for documentation, as late as possible and only when they are actually needed.
	If possible this documentation should be automatically generated from other artifacts and code.
\end{guideline}

\vspace{0.5em}
\textbf{Managing locally relevant artifacts} in continuous ways is required according to several interviewees.
Following the principle of abstraction/concreteness in Section~\ref{sec:Theoretical:Properties}, traceability to more abstract, high-level artifacts should be established.
For instance, \intB stressed this point and participated in a project for improved traceability between low-level and high-level artifacts.
\Ndx stated that both the traceability and the artifacts themselves need to be maintained:
\begin{aquote}{\ndxR}
	As many people rely on this information, especially to reason about changes, it is important that it is maintained. Today, it is up to each engineer to take care of this task.
\end{aquote}
The maintenance of information was also described as ``the tricky part'' by \ra.
For this purpose, this interviewee suggested conducting regular reviews to identify relevant artifacts and see ``which parts are less needed.''
Prescriptive models are seen as less problematic in this context, as developers are more motivated to keep artifacts updated.
We arrive at our final guideline:
\begin{guideline} \label{gui:manage_loc_rel}
	Aim to make locally relevant artifacts reusable (as with prescriptive models) and convey their relevance and use. Establish traceability to boundary objects.
\end{guideline}
\subsubsection{Summary and discussion of findings} \label{sec:rq-guidelines:discussion}
\begin{figure}
	\centering
	\includegraphics[width=0.9\linewidth]{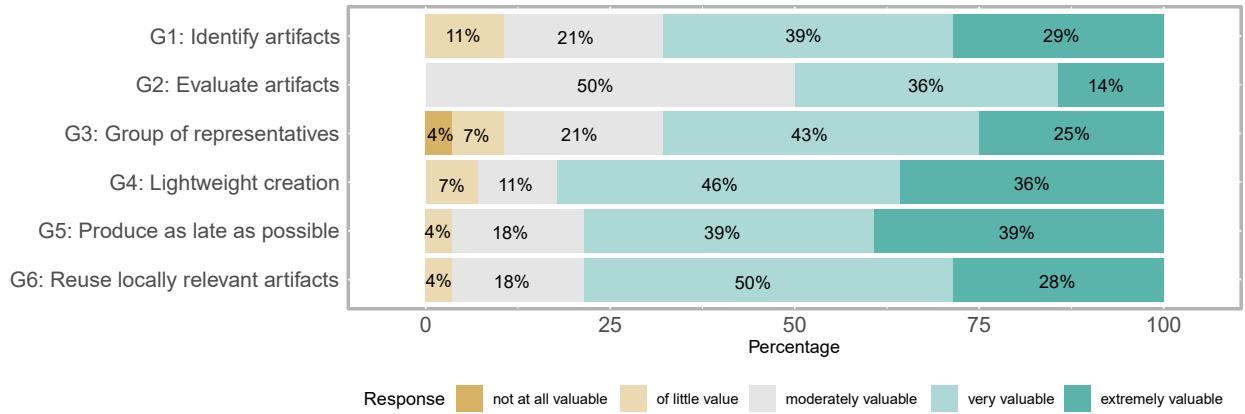}
	\caption{Questionnaire responses regarding the guidelines, n = 31}
	\label{fig:guidelines}
\end{figure}
In this section, we presented guidelines for the management of systems engineering artifacts.
These guidelines were informed by our empirical data, as well as the theoretical foundation of the area as described in Section~\ref{sec:Theoretical}.
Figure~\ref{fig:guidelines} depicts the responses to the questionnaire regarding the guidelines.
A majority of the respondents consider all guidelines at least moderately valuable.
A difference between roles could be observed with the ranking of guideline G5 (produce artifacts as late as possible): All requirements engineers and product owners considered the guideline moderately valuable or of little value, whereas 100\% of the tool support experts, testers, project/process management, and software/system architecture stakeholders ranked the guideline as very or extremely valuable.
Due to their focus, requirements engineers and product owners appear to be leaning towards early documentation and dislike creating artifacts late.
In the comments, a software developer and Scrum master added that ``boundary objects should be regularly updated'' and that ``relevant changes should be communicated, e.g., via email.''
We could not identify a big variation in the ranking of the guidelines based on the experiences of the respondents.
In the focus group sessions in Companies E and F, several participants stressed the usefulness of the distinction between boundary objects and locally relevant artifacts.

Our guidelines are based on existing guidelines and theories in the field of boundary objects.
G\ref{gui:manage_architecture} is in line with the more specialized related work on ``just enough architecting'': The upfront architecture should be minimized and all unessential decisions should be postponed until the appropriate time~\cite{Nord2006}.

There exist different approaches with respect to the frequency of updates of artifacts.
R\"{u}ping suggested to start agile documentation early in the development and update it depending on how frequently it is read, potentially even on a day-to-day basis~\cite{Ruping2003}.
We found that rather focusing on how frequently users access the artifacts, the involved actors and the importance of an artifact to create a ``common identity'' should be assessed.

It should be noted that in our model, locally relevant artifacts are always relevant within a ``team.''
Contrariwise, there can be different levels of boundary objects, shared between different companies (e.g., an OEM and its suppliers), shared in the entire company, or with a scope limited to a few teams.

\section{Boundary Objects in SystemWeaver} \label{sec:SystemWeaver}
We found that an accessible tool solution can be beneficial to maintain boundary objects.
As we elaborated on in Section~\ref{sec:Theoretical:ResidualCategories}, the way information is manifested and classified leads to the creation of standards, categories, and boundary objects.
This section comes back to the classification by focusing on SystemWeaver, a systems engineering tool used to manage information in a model-based way.
This tool allows users to collaboratively create and manage their artifacts.
In this section, we give answers to RQ3: \rqPractical

\subsection{Principles of the Tool SystemWeaver}
SystemWeaver is an industrial tool developed by the company Systemite AB.
The model-based tool is mainly used for systems engineering in the automotive domain, but not limited to it.
On a general level, SystemWeaver allows users to create \emph{items} and \emph{parts} (relations) between them.
An item can be a part of multiple items and, in this way, exist in multiple contexts.
These items and parts can be of different types that are configured in an underlying metamodel to give each type a clear semantics.
The tool is used by multiple actors who create items in relation to other items as a collaborative way of managing information.
In this way, different social groups can refer to items potentially shared between groups.
Every item has an owner and is thereby clearly connected to a social group.
The relations between items give us information about how users of information are related in a network of actors and how information is reused in different scenarios.

SystemWeaver supports the distributed development of products by allowing multiple stakeholders to share and manage required information in a synchronized and controlled way.
When data is updated, it is synchronized with the server and communicated to all clients in real time.
Groups of actors typically work on several models that are connected.
SystemWeaver supports project and change management which allow actors to be informed about changes immediately and make sure different groups' models are consistent.
Configuration management with fine-grained versioning is also a part of the tool, including mechanisms to compare versions of an item.

Item and part types can be created and managed by the end user organization in a dedicated metamodeling tool.
In the metamodeling tool, users can specify how item types are connected using relationship types.
Related information can be included, e.g., multiplicities, and attributes of item types and relationship types.

SystemWeaver provides mechanisms to traverse and visualize data in various ways.
End users can configure how graphs, grids, charts, tables, and other artifacts should be generated for various purposes.
For instance, SystemWeaver supports the import and export of AUTOSAR XML files.
\begin{figure}
	\centering
	\includegraphics[width=\linewidth]{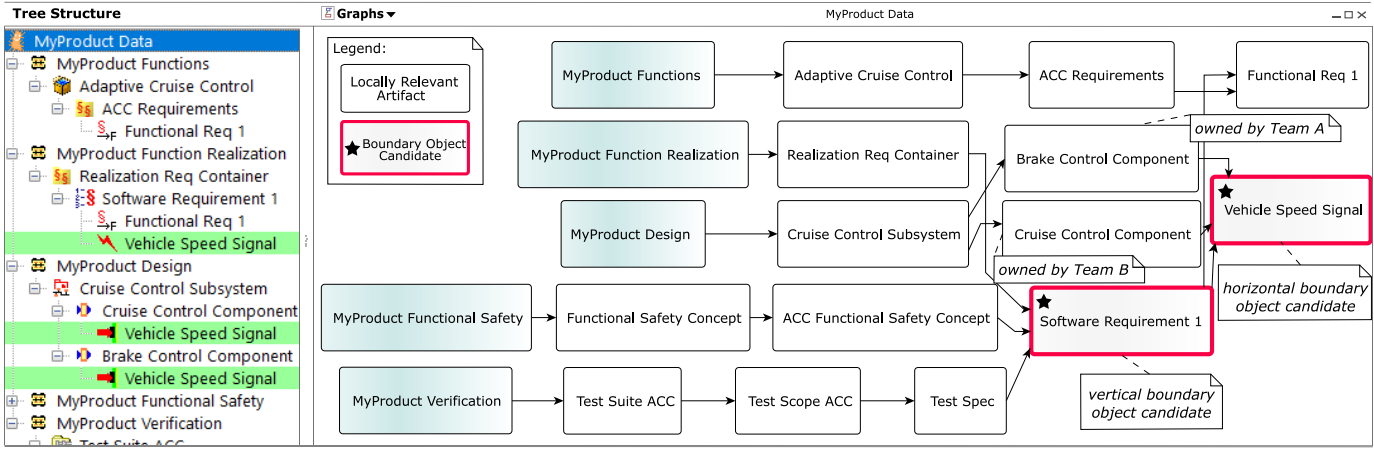}
	\caption{Excerpt of a SystemWeaver model with candidates for boundary objects}
	\label{fig:systemweaver}
\end{figure}

In Figure~\ref{fig:systemweaver}, we show an adapted screenshot of what a model looks like in SystemWeaver.
On the left side, the used artifacts are shown in a structured tree view:
The screenshot depicts the model of ``MyProduct Data'', that includes different \emph{areas} for ``MyProduct Functions'', ``MyProduct Function Realization'', ``MyProduct Design'', ``MyProduct Functional Safety'', and ``MyProduct Verification.''
Some of these areas are expanded and some are collapsed in the view.
Other items and their parts (relations) are shown in this structure.
For instance, the ``MyProduct Functions'' area contains a function for ``Adaptive Cruise Control'' that contains functional requirements.
Items can exist multiple times in this structure: The signal ``Vehicle Speed Signal'' is referred to by the ``Software Requirement 1'', in the ``Cruise Control Component'' and in the ``Brake Control Component.''
On the right side of SystemWeaver, the user can activate different visualizations of the data.
In this case, we show a configurable graph tailored to the underlying metamodel.
The boxes represent items in the model and the arrows represent the parts/relations between them.
The graph shows the areas of the example and how all underlying items are related.
Moreover, we show boundary object candidates and their types, which we will explain in more detail in Section~\ref{sec:SystemWeaver:BO}.
We added explanatory comments to the figure that were not auto-generated by the configurable graph.

SystemWeaver has been used in industrial development projects since 2003.
The evolving data has been permanently stored, which enables users to analyze the data and draw conclusions on information management concerns in systems engineering.

\subsection{Systems Engineering Artifacts in SystemWeaver} \label{sec:SystemWeaver:BO}
\begin{table*}
	\centering
	\caption{Guidelines and tool practices in SystemWeaver}
	\label{tab:Practice}
	\begin{supertabular}{>{}p{.45\textwidth}>{}p{.51\textwidth}}
		\toprule
		Guideline & {Tool practice} \tabularnewline
		\midrule
		G\ref{gui:define_strategy}: Involve stakeholders from different areas to identify what artifacts are boundary objects and locally relevant artifacts.
		& \multirow{2}{*}{\parbox{.51\textwidth}{P\ref{practice:Analyze}: Periodically analyze what items are referred to by items from different areas to identify vertical boundary objects. Analyze what items are referred to by items of different functions to identify horizontal boundary objects. Involve representatives of the main areas to validate the data.}}
		\tabularnewline	\vspace{0.5em}	
		G\ref{gui:evaluate_strategy}: Make sure that you evaluate artifacts' relevance and usage at frequent intervals. \vspace{0.5em}
		&
		\tabularnewline \midrule[0.5pt]
		G\ref{gui:manage_bos}: For each boundary object within your organization, establish a group of representatives.
		& \multirow{2}{*}{\parbox{.51\textwidth}{P\ref{practice:BoundaryObject}: Ensure that boundary objects can be stable and recognizable by releasing them at suitable intervals. Visualize them in appropriate views to stress their relations to other items. Use change notification functions and keep track of rationales for changes, e.g., using issue management features.}}
		\tabularnewline
		G\ref{gui:manage_architecture}: Find a lightweight and flexible approach to defining high-level artifacts upfront.
		&
		\tabularnewline \midrule[0.5pt]
		G\ref{gui:creation_locally_rel}: Produce locally relevant artifacts as late as possible and only when they are actually needed. Aim to generate artifacts.
		& \multirow{2}{*}{\parbox{.51\textwidth}{P\ref{practice:LocallyRelevant}: Locally relevant artifacts can have the status ``in work'' for a long period of time. Support data export and code generation features to encourage the use of prescriptive artifacts and automation. Trace links are crucial to enable P1, so exploit consistency checks to improve the trace link quality.}}
		\tabularnewline
		G\ref{gui:manage_loc_rel}: Make locally relevant artifacts reusable and convey their relevance and use. Establish traceability to boundary objects.
		& 
		\tabularnewline \bottomrule
	\end{supertabular}
\end{table*}
In the following, we discuss how to manage artifacts in SystemWeaver and suggest tool practices based on our proposed guidelines.
Table~\ref{tab:Practice} gives an overview of our guidelines and suggested tool practices.

\paragraph*{ \em Modeling artifacts in SystemWeaver}
All information artifacts in SystemWeaver are items of a certain type, which is also the case for boundary objects.
To understand which items are candidates for boundary objects, we analyzed the data of Companies C and F.
As mentioned before, we can leverage the historical data created by the users of the systems engineering tool.
We focused on the types of artifacts, relations between them, versions and releases with their timestamps, owners, and how all of these information artifacts change over time.

For our analysis, we used 11 safety-critical functions in Company C.
In our analysis, we found 9864 items in total.
When working with such a large amount of artifacts in practice, it is easy to see that stakeholders require different ways of visualizing the data.
A graph as in Figure~\ref{fig:systemweaver} is appropriate if a small subset of the data is of interest.
We created grids that displayed various interesting properties of the data.

The items are grouped in areas, related to different phases and activities of systems engineering---e.g., definition of the functions, design aspects, or verification.
The structure is based on the same metamodel as our example in Figure~\ref{fig:systemweaver}.
Stakeholders typically work in this structure in the following way: First, a function is specified with requirements, then they are broken down, then a design is made, specifying components and signals.
At the same time, functional safety analysis is made and tests are specified and executed.
It is not necessarily done in this order, but involves iterations to refine items created along the way.
As a matter of linking the different phases together, relations between items are used to refer to required information from previously created items.
In most cases, in each phase, there are different groups of stakeholders creating and owning respective items.

\paragraph*{ \em Boundary objects in SystemWeaver}
To understand the concept of boundary objects, we analyzed what items are referred to by parts from more than one main area.
After all, the creators and users of items typically have different backgrounds and tasks across areas.
It is of importance to understand how their groups interrelate and communicate, potentially using boundary objects.
We identified 827 items that were used in at least two areas.
We indicate boundary object candidates in Figure~\ref{fig:systemweaver} using a star symbol ({$\bigstar$}).

As described in Section~\ref{sec:Theoretical:Properties}, we distinguish between vertical and horizontal boundary objects.
Boundary object candidates across engineering phases are referred to as vertical, while boundary objects between development teams on the same level are called horizontal.
Of our boundary objects candidates, 351 items were of the type \emph{Software Requirement}, which are vertical boundary object candidates commonly used for several design and testing activities.
Design signals are examples of horizontal boundary object candidates, as they are pointed to by components owned by different teams who work in the same area.
In Figure~\ref{fig:systemweaver}, we indicated in a comment that the team owning the Brake Control Component is different than the team owning the Cruise Control Component.
As they share the Vehicle Speed Signal, it can be understood as a horizontal boundary object candidate.

\paragraph*{ \em Practices to manage artifacts in a systems engineering tool}
Based on our collected data and validation interviews, we understood that our presented analysis approach can support the guidelines to identify and evaluate boundary objects (Guidelines~\ref{gui:define_strategy} and \ref{gui:evaluate_strategy}).
To validate the collected data, it is recommendable to include multiple representative stakeholders in this endeavor.
We phrase this insight in the following practice:
\begin{practice} \label{practice:Analyze}
	Periodically analyze what items are referred to by items from different areas to identify vertical boundary objects. Analyze what items are referred to by items of different functions to identify horizontal boundary objects. Involve representatives of the main areas to validate the data.
\end{practice}

\begin{table*}
	\centering
	\caption{Our findings, participants, implications, relation to challenges, and discussed related work}
	\label{tab:Findings1}
	\begin{supertabular}{>{}p{.32\textwidth}>{}p{.07\textwidth}>{\raggedbottom}p{.52\textwidth}}
		\toprule
		{Finding} & {Part.} & {Implications} \tabularnewline
		\midrule
		\multicolumn{3}{l}{\textbf{Artifacts (RQ1)} --- Related work: \cite{Pareto2010,Pretschner2007, Prause2012, Selic2009, Eliasson2015}} \tabularnewline \cmidrule[0pt]{2-3}	
		{F\ref{fin:F1}: Artifacts are either important to support system-level thinking in an organization or are used inside a team.} \vspace{1.5em}
		& 1, 2, 4, 8, {E\&F}\textsuperscript{4}
		& I1: Tool solutions should allow for flexibility with the definition of important artifacts and processes to create, update, and use them. Practitioners would benefit from methods and tools supporting organization-wide collaboration. \tabularnewline 		
		{F\ref{fin:F2}: Stakeholders' motivation to manage prescriptive models is higher than dealing with descriptive artifacts.}
		& 1, 2, 5--8
		& {I2: Tools and methods supporting the automatic management of prescriptive models would be useful for practitioners. Especially descriptive models that do not play the role of boundary objects should be carefully assessed.} \tabularnewline \midrule
		\multicolumn{2}{l}{\textbf{Guidelines} --- Related work: \cite{Ruping2003,Eliasson2015,Kahkonen2004,Nord2006,Blomkvist2015,Abraham2013,Abraham2017,Levina2005}} & \textbf{Relation to Challenges (RQ2)} --- Related work: \cite{Petersen2010, Hanssen2009, Kuhrmann2018, Lagerberg2013, Ruping2003} \tabularnewline \cmidrule[0pt]{2-3}	
		G\ref{gui:define_strategy}: Involve stakeholders from different areas to identify what artifacts are boundary objects and locally relevant artifacts.
		& 2, 3, 4, 6, 10
		& I5: Reflecting on boundary objects and locally relevant artifacts helps to identify important artifacts (Ch\ref{cha:ch4}). Practitioners can establish boundary objects as a ``contract'' between plan-driven and agile teams (Ch\ref{cha:ch3}) and across locations and backgrounds (Ch\ref{cha:ch6}).
		\tabularnewline		
		G\ref{gui:evaluate_strategy}: Make sure that you evaluate artifacts' relevance and usage at frequent intervals.
		& 4, 8
		& I6: Following this guideline helps to counteract challenges with the degradation of artifacts (Ch\ref{cha:ch2}). It can also mitigate the consequences of high staff turnover (Ch\ref{cha:ch5}).
		\tabularnewline
		G\ref{gui:manage_bos}: For each boundary object within your organization, establish a group of representatives.
		& 1--7, 11
		& I7: This guideline can constitute a mechanism to align the work of different teams while allowing for diversity (Ch\ref{cha:ch1}).
		\tabularnewline
		G\ref{gui:manage_architecture}: 	Find a lightweight and flexible approach to defining high-level artifacts upfront.
		& 2, 4, 6, 8, E\&F\textsuperscript{4}
		& I8: Researchers should work towards lightweight methods to find the right level of upfront specification. Practitioners can follow the guideline to counteract artifact degradation from the start (Ch\ref{cha:ch2}).
		\tabularnewline \cmidrule[0pt]{2-3}
		G\ref{gui:creation_locally_rel}: Produce locally relevant artifacts  as late as possible and only when they are actually needed. Aim to generate artifacts.
		& 3, 5, 8
		& I9: Producing artifacts as late as possible helps to avoid the degradation of artifacts (Ch\ref{cha:ch2}). This guideline is in line with F\ref{fin:F2} as the motivation to manage prescriptive models (for artifact generation) was found to be high.
		\tabularnewline
		G\ref{gui:manage_loc_rel}: Make locally relevant artifacts reusable and convey their relevance and use. Establish traceability to boundary objects.
		& 2, 7, 8, 9
		& I10: Practitioners trying to align different groups in the organization (Ch\ref{cha:ch1}) should focus on traceability to boundary objects. Researchers can focus on increasing the reusability of artifacts to alleviate the maintenance effort.
		\tabularnewline \bottomrule
	\end{supertabular}
	\textsuperscript{4} E\&F means that there existed a general consensus about the finding/guideline in the focus groups.
\end{table*}

It is noteworthy that there might exist boundary objects that are not directly referred to by an item, but affect their content.
A tool expert stated that the architecture model containing components and their structure plays the role of a horizontal boundary object, as it provides an abstraction of the system and clarifies the relations between development teams.
In that way, containing items for each of the main areas can also be candidates for boundary objects (e.g., architecture model).

When analyzing the time aspect, it can be noted that vertical boundary objects are typically created in a top-down approach, starting with the function specification that is refined, allocated to systems, and verified.
The creation dates of the items containing vertical boundary objects differ by approximately 1.5 years in our example model.
In some cases, items are reused throughout generations of platforms and are referred to by even more contexts over many years.
In our example, horizontal boundary objects (e.g., signals between components) are typically referred to by other items within approximately 4-5 months.

It should be noted that an important property of boundary objects is that they are actual ``boundary objects-in-use''~\cite{Levina2005}.
For this reason, we analyzed release processes and differences in the change management of different boundary objects.
In fact, all boundary object candidates were released and 705 of them had a version number higher than 1.
Also our empirical findings and theoretical foundation point to the importance of stability of boundary objects and are in line with the finding that boundary objects should be released and versioned.
The importance of change management is key to ensure that boundary objects can be leveraged for collaboration.
Moreover, appropriate visualization mechanisms facilitate the traversal of data and the identification of important artifacts, as we described in Section~\ref{sec:Theoretical:Properties} and~\ref{sec:RQChallenges}.
We suggest the following practice to manage boundary objects (Guidelines~G\ref{gui:manage_bos} and G\ref{gui:manage_architecture}):

\begin{practice}  \label{practice:BoundaryObject}
	Ensure that boundary objects can be stable and recognizable by releasing them at suitable intervals. Visualize them in appropriate views to stress their relations to other items. Use change notification functions and keep track of rationales for changes, e.g., using issue management features.
\end{practice}

We saw that more of the locally relevant artifacts had the status ``in work'' and are not released as stringently.
In fact, also our guidelines G\ref{gui:creation_locally_rel} and G\ref{gui:manage_loc_rel} suggest to produce and manage locally relevant artifacts more flexibly.
It is an advantage if they can be reused for different products over time.
As developers are more motivated to maintain prescriptive artifacts, respective automation features should be supported and improved.
Trace links and relations between items have a great potential for the analysis of artifacts and should be invested in to support P\ref{practice:Analyze}.
We capture these findings in the following practice:
\begin{practice}  \label{practice:LocallyRelevant}
	Locally relevant artifacts can have the status ``in work'' for a long period of time. Support data export and code generation features to encourage the use of prescriptive artifacts and automation. Trace links are crucial to enable P1, so exploit consistency checks to improve the trace link quality.
\end{practice}
\section{Concluding Remarks and Future Work} \label{sec:Discussion}
In this paper, we presented empirical findings and guidelines for the management of systems engineering artifacts in agile automotive contexts.
We based this study on a design-science approach with more than 50 practitioners from six companies.
We used theoretical concepts from the field of actor-network theory and boundary objects to inform this study.
To understand the practical use of boundary objects, we analyzed the development data of two automotive companies and describe practices to manage artifacts.

In Table~\ref{tab:Findings1}, an overview of our findings with implications for research and practice is provided.
We also show the relation between the guidelines and the identified challenges (RQ2) in the Implications column.
The leftmost column shows references to related work discussed in Section~\ref{sec:Finding}.
The participants (see Table~\ref{tab:Participants}) who mainly support the findings are shown in the column ``Part.''

Our findings suggest that it is central to identify artifacts and evaluate artifacts' relevance and usage at frequent intervals.
Boundary objects (used to create a shared understanding between several actors) should be managed upfront with a lightweight approach and be continuously revised.
Artifacts used within one team can be managed with a flexible approach and should only be created when they are actually needed.

Our guidelines and practices help practitioners to understand the use and relevance of artifacts and how to manage them, potentially supported by a tool.
Researchers can benefit from a new perspective on continuous management of systems engineering artifacts based on empirical evidence and a theoretical foundation based on actor-network and boundary object theory.

\noindent {\bf Future work}:
We expect that our contributions can be the basis for further, more tailored practices for specific concerns (e.g., architecture).
Moreover, actual boundary objects and locally relevant artifacts might differ depending on the involved disciplines and relationships of involved companies (e.g., supplier and OEM relationships).
There are particular characteristics of the domain under study that need to be taken into consideration.
We believe that the presented guidelines are valuable also for other domains, but dedicated studies are needed to scrutinize their applicability.

\section*{Acknowledgments}
We are very grateful for the support of the participants in this study and we thank
for all the clarifications provided when needed.

This work was partially supported by the Next Generation Electrical Architecture (NGEA) and NGEA step 2 projects by VINNOVA FFI (2014-05599 and 2015-04881), the Software Center Project 27 on RE for Large-Scale Agile System Development, and by the Wallenberg AI, Autonomous Systems and Software Program (WASP) funded by the Knut and Alice Wallenberg Foundation.


\begin{thebibliography}{10}
	
	\bibitem{Dyba2008}
	Dyb{\aa} Tore, Dings{\o}yr Torgeir. Empirical studies of agile software
	development: A systematic review.  {\it Information and Software Technology.
	}2008;50(9-10):833--859.
	
	\bibitem{Houston2014}
	Houston Dan~X.. Agility beyond software development.  In: Proceedings of the
	International Conference on Software and System Process (ICSSP'14):65--69ACM;
	2014; New York, NY, USA.
	
	\bibitem{Turner2007}
	Turner Richard. Toward Agile systems engineering processes.  {\it Cross Talk,
		The Journal of Defense Software Engineering. }2007;.
	
	\bibitem{Haberfellner2005}
	Haberfellner Reinhard, Weck Olivier. 10.1.3 Agile SYSTEMS ENGINEERING versus
	AGILE SYSTEMS engineering.  {\it INCOSE International Symposium.
	}2005;15(1):1449-1465.
	
	\bibitem{Dingsoyr2017a}
	Dings{\o}yr Torgeir, Moe Nils~Brede, Faegri Tor~Erlend, Seim Eva~Amdahl.
	Exploring software development at the very large-scale: a revelatory case
	study and research agenda for agile method adaptation.  {\it Empirical
		Software Engineering. }2017;.
	
	\bibitem{Ebert2017}
	Ebert Christof, Favaro John. Automotive Software.  {\it IEEE Software.
	}2017;34(3):33--39.
	
	\bibitem{DAmbrosio2017}
	D'Ambrosio Joseph, Soremekun Grant. Systems engineering challenges and {MBSE}
	opportunities for automotive system design.  In: International Conference on
	Systems, Man, and Cybernetics (SMC):2075--2080IEEE; 2017.
	
	\bibitem{Pareto2010}
	Pareto Lars, Eriksson Peter, Ehnebom Staffan. Architectural Descriptions as
	Boundary Objects.  In: Proceedings of the 13th International Conference on
	Model Driven Engineering Languages and Systems (MODELS 2010):406--419Springer
	Berlin Heidelberg; 2010; Berlin, Heidelberg.
	
	\bibitem{Hummel2013}
	Hummel Markus, Rosenkranz Christoph, Holten Roland. The role of communication
	in agile systems development: An analysis of the state of the art.  {\it
		Business and Information Systems Engineering. }2013;5(5):343--355.
	
	\bibitem{Ruping2003}
	R{\"{u}}ping Andreas. {\it Agile Documentation: A Pattern Guide to Producing
		Lightweight Documents for Software Projects}.
	\newblock Wiley Publishing; 1~ed.2003.
	
	\bibitem{Dingsoyr2017}
	Dings{\o}yr Torgeir, Moe Nils~Brede, Faegri Tor~Erlend, Seim Eva~Amdahl.
	Exploring software development at the very large-scale: a revelatory case
	study and research agenda for agile method adaptation.  {\it Empirical
		Software Engineering. }2018;23(1):490--520.
	
	\bibitem{Kajko-Mattsson2008}
	Kajko-Mattsson Mira. Problems in agile trenches.  In: Proceedings of the 2nd
	International Symposium on Empirical Software Engineering and Measurement
	(ESEM'08)ACM; 2008; New York, NY, USA.
	
	\bibitem{Scacchi1984}
	Scacchi Walt. Managing software engineering projects: A social analysis.  {\it
		IEEE Transactions on Software Engineering. }1984;(1):49--59.
	
	\bibitem{Star1989}
	Star Susan~Leigh, Griesemer James~R.. Institutional Ecology, `Translations' and
	Boundary Objects: Amateurs and Professionals in Berkeley's Museum of
	Vertebrate Zoology, 1907-39.  {\it Social Studies of Science.
	}1989;19(3):387--420.
	
	\bibitem{Wohlrab2018}
	Wohlrab Rebekka, Pelliccione Patrizio, Knauss Eric, Larsson Mats. Boundary
	Objects in Agile Practices: Continuous Management of Systems Engineering
	Artifacts in the Automotive Domain.  In:  Kuhrmann Marco, O'Connor Rory~V.,
	Houston Dan, eds. {\it Proceedings of the 2018 International Conference on
		Software and System Process, {ICSSP} 2018, Gothenburg, Sweden, May 26-27,
		2018}, :31--40{ACM}; 2018.
	
	\bibitem{Hevner2004}
	Hevner Alan~R., March Salvatore~T., Park Jinsoo, Ram Sudha. Design Science in
	Information Systems Research.  {\it MIS Quarterly. }2004;28:75--105.
	
	\bibitem{Petersen2010}
	Petersen Kai, Wohlin Claes. The effect of moving from a plan-driven to an
	incremental software development approach with agile practices: An industrial
	case study.  {\it Empirical Software Engineering. }2010;15(6):654--693.
	
	\bibitem{Lagerberg2013}
	Lagerberg Lina, Skude Tor, Emanuelsson Par, Sandahl Kristian, Stahl Daniel. The
	impact of agile principles and practices on large-scale software development
	projects: A multiple-case study of two projects at {Ericsson}.  In:
	International Symposium on Empirical Software Engineering and
	Measurement:348--356IEEE; 2013.
	
	\bibitem{Kuhrmann2013}
	Kuhrmann Marco, {M{\'e}ndez Fern{\'a}ndez} Daniel, Gr{\"o}ber Matthias. Towards
	Artifact Models as Process Interfaces in Distributed Software Projects.  In:
	2013 IEEE 8th International Conference on Global Software Engineering:11-20;
	2013.
	
	\bibitem{Mendez2010}
	M{\'e}ndez~Fern{\'a}ndez Daniel, Penzenstadler Birgit, Kuhrmann Marco, Broy
	Manfred. A Meta Model for Artefact-Orientation: Fundamentals and Lessons
	Learned in Requirements Engineering.  In:  Petriu Dorina~C., Rouquette
	Nicolas, Haugen {\O}ystein, eds. {\it Model Driven Engineering Languages and
		Systems}, :183--197Springer Berlin Heidelberg; 2010; Berlin, Heidelberg.
	
	\bibitem{Mendez2018}
	M{\'e}ndez~Fern{\'a}ndez Daniel, B{\"o}hm Wolfgang, Vogelsang Andreas, et al.
	Artefacts in Software Engineering: What are they after all?.  {\it arXiv
		preprint arXiv:1806.00098. }2018;.
	
	\bibitem{Bass2016}
	Bass Julian. Artefacts and agile method tailoring in large-scale offshore
	software development programmes.  {\it Information and Software Technology.
	}2016;75:1--16.
	
	\bibitem{Voigt2016}
	Voigt Stefan, Garrel J\"{o}rg, M\"{u}ller Julia, Wirth Dominic. A Study of
	Documentation in Agile Software Projects.  In: Proceedings of the 10th
	ACM/IEEE International Symposium on Empirical Software Engineering and
	Measurement:4:1--4:6ACM; 2016; New York, NY, USA.
	
	\bibitem{Heldal2016}
	Heldal Rogardt, Pelliccione Patrizio, Eliasson Ulf, Lantz Jonn, Derehag Jesper,
	Whittle Jon. Descriptive vs Prescriptive Models in Industry.  In: Proceedings
	of the 19th International Conference on Model Driven Engineering Languages
	and Systems (MODELS 2016):216--226ACM; 2016; New York, NY, USA.
	
	\bibitem{Malone1994}
	Malone Thomas~W., Crowston Kevin. The Interdisciplinary Study of Coordination.
	{\it ACM Computing Surveys. }1994;26(1):87--119.
	
	\bibitem{Callon1986}
	Callon Michel. Some elements of a sociology of translation: Domestication of
	the scallops and the fishermen of St Brieuc Bay.  In:  Law John, ed. {\it
		Power, Action and Belief: A New Sociology of Knowledge}, London: Routledge \&
	Kegan Paul 1986.
	
	\bibitem{Latour1987}
	Latour Bruno. {\it Science in action: How to follow scientists and engineers
		through society}.
	\newblock Harvard University Press; 1987.
	
	\bibitem{Blomkvist2015}
	Blomkvist Johan~Kaj, Persson Johan, {\AA}berg Johan. Communication through
	Boundary Objects in Distributed Agile Teams.  In: Proceedings of the 33rd
	Annual ACM Conference on Human Factors in Computing Systems
	(CHI'15):1875--1884ACM; 2015; New York, NY, USA.
	
	\bibitem{Bechky2003}
	Bechky Beth~A.. Sharing Meaning Across Occupational Communities: The
	Transformation of Understanding on a Production Floor.  {\it Organization
		Science. }2003;14(3):312-330.
	
	\bibitem{Wohlrab2016}
	Wohlrab Rebekka, Stegh{\"{o}}fer Jan-Philipp, Knauss Eric, Maro Salome, Anjorin
	Anthony. Collaborative Traceability Management: Challenges and Opportunities.
	In: 24th IEEE International Requirements Engineering Conference
	(RE'16):216--225; 2016.
	
	\bibitem{Maro2018}
	Maro Salome, Stegh{\"{o}}fer Jan-Philipp, Staron Miroslaw. Software
	traceability in the automotive domain: Challenges and solutions.  {\it
		Journal of Systems and Software. }2018;141:85 - 110.
	
	\bibitem{Pretschner2007}
	Pretschner Alexander, Broy Manfred, Kr{\"{u}}ger Ingolf~H., Stauner Thomas.
	Software engineering for automotive systems: A roadmap.  In: Future of
	Software Engineering (FoSE'07):55--71IEEE; 2007.
	
	\bibitem{Pelliccione2017}
	Pelliccione Patrizio, Knauss Eric, Heldal Rogardt, et al. Automotive
	Architecture Framework: The Experience of Volvo Cars.  {\it Journal of
		Systems Architecture. }2017;77.
	
	\bibitem{Eliasson2015RE}
	Eliasson Ulf, Heldal Rogardt, Knauss Eric, Pelliccione Patrizio. The need of
	complementing plan-driven requirements engineering with emerging
	communication: Experiences from Volvo Car Group.  In: Proceedings of the IEEE
	23rd International Requirements Engineering Conference (RE'15):372--381IEEE;
	2015.
	
	\bibitem{Wohlrab2018REFSQ}
	Wohlrab Rebekka, Pelliccione Patrizio, Knauss Eric, Gregory Sarah~C.. The
	Problem of Consolidating RE Practices at Scale: An Ethnographic Study.  In:
	Requirements Engineering: Foundation for Software Quality:155--170Springer
	International Publishing; 2018; Cham.
	
	\bibitem{ISO26262}
	{International Organization for Standardization} . Road vehicles -- Functional
	safety.  {\it ISO26262:2011. }2011;.
	
	\bibitem{Automotive2017}
	{VDA QMC Working Group 13 / Automotive SIG} . Automotive SPICE Process
	Assessment / Reference Model.  {\it Automotive SPICE:2017. }2017;3.1.
	
	\bibitem{Diebold2017}
	Diebold Philipp, Zehler Thomas, Richter Dominik. How do agile practices support
	automotive {SPICE} compliance?.  In: Proceedings of the International
	Conference on Software and System Process (ICSSP'17):80--84ACM; 2017; New
	York, NY, USA.
	
	\bibitem{Abraham2017}
	Abraham Ralf. Guidelines for Architecture Models as Boundary Objects:193--210.
	\newblock Cham: Springer International Publishing 2017.
	
	\bibitem{Bowker1999}
	Bowker Geoffrey~C., Star Susan~Leigh. {\it Sorting things out: classification
		and its consequences}.
	\newblock Cambridge, Mass.: MIT Press; 1999.
	
	\bibitem{Abraham2013}
	Abraham Ralf. Enterprise Architecture Artifacts as Boundary Objects - A
	Framework of Properties.  In: Proceedings of the 21st European Conference on
	Information Systems (ECIS 2013):120Association for Information Systems; 2013.
	
	\bibitem{Lave1991}
	Lave Jean. Situating learning in communities of practice.  {\it Perspectives on
		socially shared cognition. }1991;2:63--82.
	
	\bibitem{Wenger2002}
	Wenger Etienne, McDermott Richard~Arnold, Snyder William. {\it Cultivating
		communities of practice: A guide to managing knowledge}.
	\newblock Harvard Business Press; 2002.
	
	\bibitem{Levina2005}
	Levina Natalia, Vaast Emmanuelle. The Emergence of Boundary Spanning Competence
	in Practice: Implications for Implementation and Use of Information Systems.
	{\it MIS Quarterly. }2005;29(2):335--363.
	
	\bibitem{Creswell2008}
	Creswell John~W.. {\it Research Design: Qualitative, Quantitative, and Mixed
		Methods Approaches}.
	\newblock London, England: Sage Publications Ltd.; 3~ed.2008.
	
	\bibitem{Runeson2009}
	Runeson Per, H{\"{o}}st Martin. Guidelines for conducting and reporting case
	study research in software engineering.  {\it Empirical Software Engineering.
	}2009;14(2):131--164.
	
	\bibitem{Tremblay2010}
	Tremblay Monica~Chiarini, Hevner Alan~R., Berndt Donald~J.. The Use of Focus
	Groups in Design Science Research.  In: Integrated Series in Information
	Systems. Springer US 2010 (pp. 121--143).
	
	\bibitem{Likert1932}
	Likert Rensis. A technique for the measurement of attitudes.  {\it Archives of
		Psychology. }1932;22(140):5--55.
	
	\bibitem{Atkinson2001}
	Atkinson Rowland, Flint John. Accessing hidden and hard-to-reach populations:
	Snowball research strategies.  {\it Social research update. }2001;33(1):1--4.
	
	\bibitem{Leffingwell2007}
	Leffingwell Dean. {\it Scaling Software Agility: Best Practices for Large
		Enterprises (The Agile Software Development Series)}.
	\newblock Addison-Wesley Professional; 2007.
	
	\bibitem{Schwaber2001}
	Schwaber Ken, Beedle Mike. {\it Agile Software Development with Scrum}.
	\newblock Prentice Hall PTR; 1~ed.2001.
	
	\bibitem{maxwell2012qualitative}
	Maxwell Joseph~Alex. {\it Qualitative Research Design: An Interactive
		Approach}.
	\newblock Applied Social Research MethodsSAGE Publications; 2012.
	
	\bibitem{Prause2012}
	Prause Christian~R., Durdik Zoya. Architectural design and documentation: Waste
	in agile development?.  In: Proceedings of the International Conference on
	Software and System Process (ICSSP'12):130--134IEEE; 2012.
	
	\bibitem{Selic2009}
	Selic Bran. Agile documentation, anyone?.  {\it IEEE Software.
	}2009;26(6):11--12.
	
	\bibitem{Eliasson2015}
	Eliasson Ulf, Heldal Rogardt, Pelliccione Patrizio, Lantz Jonn. Architecting in
	the Automotive Domain: Descriptive vs Prescriptive Architecture.  In:
	Proceedings of the 12th Working IEEE/IFIP Conference on Software Architecture
	(WICSA'15):115--118IEEE; 2015.
	
	\bibitem{Carlile2002}
	Carlile Paul~R.. A Pragmatic View of Knowledge and Boundaries: Boundary Objects
	in New Product Development.  {\it Organization Science. }2002;13(4):442--455.
	
	\bibitem{Hanssen2009}
	Hanssen Geir~K., Yamashita Aiko~Fallas, Conradi Reidar, Moonen Leon.
	Maintenance and agile development: Challenges, opportunities and future
	directions.  In: Proceedings of the 25th IEEE International Conference on
	Software Maintenance (ICSM'09):487--490IEEE; 2009.
	
	\bibitem{Kuhrmann2018}
	Kuhrmann Marco, Diebold Philipp, M\"unch J\"urgen, et al. Hybrid Software
	Development Approaches in Practice: A European Perspective.  {\it IEEE
		Software. }2018;PP(99).
	
	\bibitem{Nord2006}
	Nord Robert~L., Tomayko James~E.. Software architecture-centric methods and
	agile development.  {\it IEEE Software. }2006;23(2):47-53.
	
	\bibitem{Kahkonen2004}
	K{\"{a}}hk{\"{o}}nen Tuomo. Agile Methods for Large Organizations -- Building
	Communities of Practice.  In: Proceedings of the Agile Development
	Conference:2--10IEEE; 2004.
	
\end{thebibliography}
\end{document}